\documentclass{article}
\usepackage{indentfirst} % add indentation to the first par after a heading
\usepackage{graphicx} % package to manage images
\usepackage{float} % required by fixing img position by \begin{figure}[H]
\usepackage{amsmath} % required by matrices and equations
\usepackage{bm} % bold greek letters

\usepackage{authblk} % to add affiliation

\usepackage[skip=2pt,font=scriptsize]{caption}
\usepackage{subcaption} % required by subfigure
\usepackage{geometry} % page geometry
\usepackage{setspace} % required by line spacing adjustment
\usepackage{textcomp, gensymb} % required by generating degree symbol. textcomp removes warnings from gensymb.
\usepackage{lineno} % required by linenumber
\usepackage{textgreek} % required by typing greek letter in text
\usepackage{tabularx} % required to control table width
\usepackage{threeparttable} % to use table notes
\usepackage{makecell} % to change line in table
\usepackage{pdflscape} % to rotate table
\usepackage{setspace}

\usepackage{natbib} %Import the natbib package and sets a bibliography  and citation styles
\title{\vspace{-5ex}\textbf{Microstructural evolution of Carrara marble during semi-brittle deformation}\vspace{-1ex}}

\author[1]{Tongzhang Qu}
\author[1,2]{Nicolas Brantut}
\author[3]{David Wallis}
\author[1,4]{Christopher Harbord}
\affil[1]{Department of Earth Sciences, University College London, London, UK}
\affil[2]{GFZ German Center for Geosciences, Potsdam, Germany}
\affil[3]{Department of Earth Sciences, University of Cambridge, Cambridge, UK}
\affil[4]{Galson Sciences Ltd., Oakham, UK}
\date{}

\begin{document}
\maketitle
\vspace{-5ex}Corresponding author: Tongzhang Qu (tongzhang.qu@ucl.ac.uk)
\section*{Key points}
    \begin{itemize}
        \setlength\itemsep{0.01em}
        \item With axial strain up to 2\%, hardening is high, twins accommodate most axial deformation, and fracture intensity increases rapidly in marble.
        \item Beyond 2\% strain, hardening decreases, twins and cracks accumulate less rapidly, and geometrically necessary dislocation density increases.
        \item Yield stress and overall strength depend strongly on temperature, indicating dislocation glide as strength limiting process.
    \end{itemize}

\section*{Abstract}
    Fifteen marble samples were subjected to semi-brittle deformation through triaxial compression experiments, reaching axial strains of 0.5\%, 1.0\%, 2.0\%, 4.0\%, or 7.5\% at temperatures of 20\textdegree{}C, 200\textdegree{}C, or 350\textdegree{}C, under a confining pressure of 400 MPa. Deformation twins, lattice curvature, and intragranular microfractures in the samples were quantitatively characterised using forescattered electron images and electron backscatter diffraction. Microstructural analyses revealed that twins accommodate most of the shortening during the first 2\% strain, whereas lattice curvature associated with geometrically necessary dislocations predominantly develops in the later stages. Intragranular fracture intensity exhibits an almost linear correlation with strain during the first 2\% strain but increases more slowly thereafter. The mechanical data indicate a strong temperature dependence of yield stress, consistent with the temperature dependence of the critical resolved shear stress for dislocation glide. The subsequent strain hardening is likely caused by progressively increasing intensity of interactions among dislocations and between dislocations and twin boundaries. Based on the microstructural data and interpreted hardening mechanisms, we propose a phenomenological model, with microstructural state variables, for semi-brittle deformation at our experimental conditions as a step towards development of a microphysical constitutive model of semi-brittle deformation.

\section*{Plain language summary}
    %\singlespace{}
    In the shallow crust, rocks deform by fracturing and faulting. With increasing depth, elevated pressure and temperature make rocks more ductile. The transition between the shallow, brittle regime and the deeper ductile regime is where rocks are the strongest across the lithosphere. However, it is not clear what controls this strength. In particular, the role of several deformation mechanisms (fracturing and plastic deformation) and how they interact is not known. In this study, we simulate deformation at intermediate conditions between the brittle and ductile regime in the laboratory, and quantify how deformation mechanisms interact at different deformation stages. To achieve this, we squeezed samples of calcite marble at elevated pressure and a range of temperature, and we measured the extent of microcracking and plastic deformation activity using electron microscopy techniques. Our work shows that different mechanisms are activated in sequence, with a lot of cracking coupled to plastic twinning at low strain, and then a substantial increase in dislocation activity at large strain. We find that the overall strength is most likely controlled by dislocation activity, which is the hardest deformation mechanism in the conditions tested. We suggest that our dataset can be used to develop a rheological model for rocks across the brittle-ductile transition. %the extent of flowing and fracturing was measured as different quantities from analysis of images and other information obtained from microscope. From our analysis, we have a better understanding of the processes of semi-brittle deformation and how the strength of marble changes with semi-brittle deformation. 

\section{Introduction}
    As pressure and temperature increase with depth, the mechanical behaviour of the lithosphere undergoes a transition from primarily pressure-dependent frictional processes and associated brittle behaviour to primarily temperature-dependent crystal-plastic deformation \citep{Brace_1980, Kohlstedt_1995}. At depths spanning this transition, the concurrent operation of brittle and crystal-plastic processes, referred to as semi-brittle deformation, is evidenced by a broad range of field observations from outcrops of exhumed mid-crustal rocks \citep[e.g.,][]{Bak_1975, Sibson_1977, Passchier_1982, white_1983, Fagereng_2010}. Understanding of semi-brittle deformation is of fundamental importance to modelling deformation of the lithosphere as the semi-brittle regime is where the strength of rocks is greatest. Thus, semi-brittle deformation exerts a key control on several aspects of geodynamics on Earth, including integrated plate strength \citep{Kohlstedt_1995}, the depth extent of the seismogenic zone \citep{Sibson_1982,Shimamoto_1986,Carpenter_2016}, and characteristics of lithospheric flexure \citep{Chapple_1979, Sandiford_2023}. \par 
    
    Quantitative characterisation of the complex interplay of deformation mechanisms operating during semi-brittle deformation requires combined efforts from laboratory experimentation, microstructural observation, and micromechanical modelling. Calcite aggregates have been frequently chosen as the experimental material to study semi-brittle deformation \citep[e.g.,][]{Olsson_1974, Fredrich_1989, Rybacki_2021, Harbord_2023}, not only because calcite is an important rock-forming mineral, but also because the pressures and temperatures required for semi-brittle deformation of calcite aggregates are more accessible in experiments than those required for semi-brittle deformation of silicates. At the macroscopic scale, semi-brittle deformation of calcite aggregates is characterised by strain-hardening flow with an absence of localised failure \citep[e.g.,][]{Fredrich_1989}. Qualitative microstructural characterisation of marble samples tested in the semi-brittle regime has revealed concurrent operation of twinning, dislocation activity, and microfracturing \cite[][and references therein]{Olsson_1976, Carter_1978, Fredrich_1989, Rybacki_2021}. Correspondingly, quantitative measurements have been made of three main microstructural parameters, specifically twin density \citep[e.g.,][]{Rutter_1983, Rybacki_2013, Rutter_2022}, dislocation density \citep[e.g.,][]{Fredrich_1989, de_1996}, and fracture density/intensity \citep[e.g.,][]{Fredrich_1989, Harbord_2023}. \par
    
    Despite these efforts, a robust constitutive model for semi-brittle deformation remains elusive for two main reasons. First, the evolution of microstructures with macroscopic strain has not been well captured. This limitation is because most previous work focused on the microstructures of marble at similar strains but different pressure-temperature conditions \citep{Fredrich_1989, Rybacki_2021, Harbord_2023}, lacking a systematic investigation of microstructures at different strains under the same experimental conditions. Due to the interplay between microfractures and other crystal defects, the evolution of microstructures with macroscopic strain cannot be predicted quantitatively based on existing knowledge of purely frictional or crystal-plastic deformation. Second, there is no consensus on the relationships among mechanical properties and microstructures during semi-brittle deformation. Numerous hardening mechanisms can be activated, including increasing the number of obstacles to dislocation glide \citep{Rybacki_2021}, increasing resistance to frictional sliding of microfractures under confinement \citep{Walsh_1965}, and increasing intergranular back stress at twin tips \citep{Burkhard_1993}. The relative importance of each mechanism is currently unclear. As such, quantitative rheological models for semi-brittle flow are currently limited and rely on strongly simplifying assumptions and poorly constrained quantities \citep{Horii_1986, Nicolas_2017, Liu_2023}.\par 
     
    Given these limitations of previous studies, we aim to provide quantitative constraints on the development of brittle and crystal-plastic microstructures in marble undergoing semi-brittle deformation. We conducted triaxial compression experiments on samples of Carrara marble up to axial strains of 0.5\%, 1.0\%, 2.0\%, 4.0\% or 7.5\% at temperatures of 20\textdegree{}C, 200\textdegree{}C or 350\textdegree{}C and a confining pressure of 400 MPa. The post-mortem samples deformed to different strains were characterised by forescattered electron (FSE) images and electron backscatter diffraction (EBSD). Comprehensive microstructural information, including twin density, twin spacing, geometrically necessary dislocation density, and fracture intensity were collected as a function of macroscopic strain and temperature. We relate the mechanical data to the microstructural data by interpreting the micromechanical mechanisms for the onset of inelasticity and subsequent strain hardening. Based on our interpretations, we propose a phenomenological model for semi-brittle deformation with the collected microstructural quantities as state variables to serve as a guide to the development of future microphysical constitutive models.\par

\section{Methods}

\subsection{Experimental materials}
    The experimental material for this study was Carrara marble. This marble is composed of polycrystalline calcite (\textgreater{} 99\%), with \textless{} 1\% porosity and no crystallographic preferred orientation. Most grains in Carrara marble are free of twins and other deformation microstructures. A small proportion of grains contain pre-existing thick twins, $\geq$ 5 \textmu m in width, which are distinguishable from the thin deformation twins, $\leq$ 2 \textmu m in width, that are typically imparted during deformation experiments. Cylindrical samples of 9.8 mm in diameter were drilled from a block of Carrara marble and ground to a length of 22 mm. After cleaning with water, the samples were dried in an oven at 70 \textdegree C for at least 30 hours before mechanical testing. \par

\subsection{Mechanical testing}
    \label{sec:methods_mechanical_testing}
    We conduced triaxial deformation experiments in the recently refurbished Murrell gas-medium apparatus at University College London \citep{Edmond_1973, Murrell_1989, Harbord_2022}. Details of the refurbishment are provided by \citet{Harbord_2022}. Before deformation experiments, each cylindrical sample was sandwiched between two Inconel disks, each  with 10 mm in diameter and 5 mm in height. The sample and Inconel disks were loaded into an annealed copper jacket that was 0.2 mm thick and 50 mm in length with an inner diameter of 10.5 mm. Two Inconel pistons were inserted into the two ends of the copper jacket loaded with the sample and disks. Two Inconel rings were swaged over the contact between the copper jacket and the Inconel pistons. To limit overstretching and avoid puncture of the jacket, a small amount of molybdenum disulfide was applied to the outer surface of the copper jacket before swaging. The resulting collapse of the copper jacket onto the Inconel pistons provided a gas seal for the sample during mechanical testing. \par
    
\begin{figure}[H]
        \centering
        \includegraphics[width=0.8\textwidth]{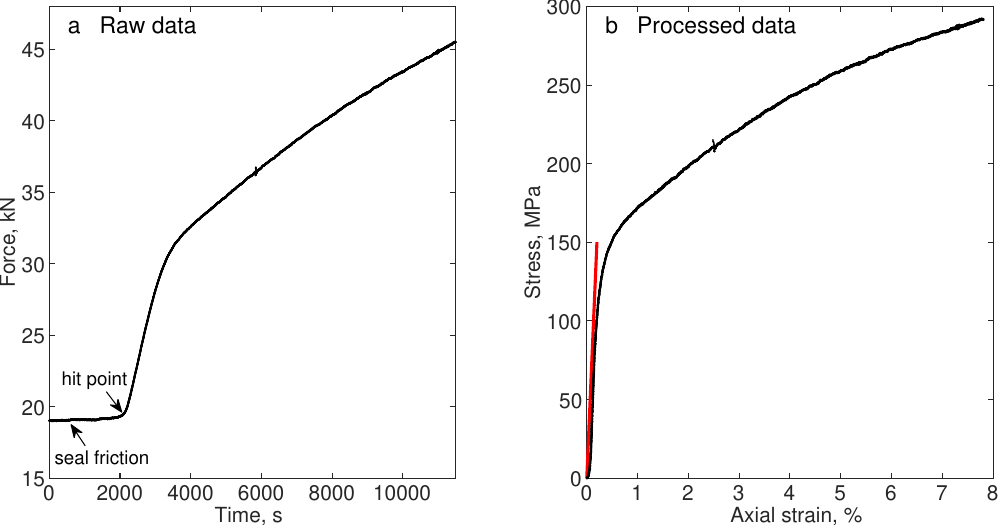}
        \caption{(a) Raw data from Run 244 at a temperature of 350\textdegree{}C and confining pressure of 400 MPa. (b) Processed stress-strain data from Run 244 in black. The red line is estimated stress from the Voigt-Reuss-Hill average Young's modulus ($E_{\mathrm{0}}$) at the same pressure and temperature for reference. The raw data and stress-strain data in this study are plotted as dots. The apparent thicknesses of the curves reflect the scatter of mechanical data.}
        \label{fig:data_processing}
        \end{figure}

    The specimen assembly was screwed on a steel piston and loaded through a series of Bridgeman seals into the deformation apparatus \citep{Edmond_1973, Murrell_1989}. High confining pressure was achieved by pumping Ar into the pressure vessel with Bridgeman seals between the vessel liner and the steel piston. The top of the steel piston was located outside of the pressure vessel. Elevated temperatures at the sample were achieved by a furnace within the pressure vessel and calibrated at the target temperatures before mechanical experiments. The differential axial force was applied by a servo hydraulic actuator located vertically above the pressure vessel and transmitted through an external load cell onto the top of the steel piston. The precision of force measurement is on the order of 0.1 kN, indicated by the scatter of raw data in Figure \ref{fig:data_processing}a. The shortening of the specimen assembly was measured by a pair of RDP GT5000-L25 linear variable differential transducers, with a precision of 150 nm, attached to the top steel piston \citep[Figure 2 of][]{Harbord_2022}.\par
    
    The raw data of axial force and displacement, both measured externally from the pressure vessel, required corrections to obtain stress and strain of the sample (Figure \ref{fig:data_processing}). First, the seal friction and strength of the copper jacket were subtracted from the raw force data. The seal friction was found to be dependent on pressure and displacement rate. Here, a constant confining pressure of 400 MPa and displacement rate of 220 nm/s, corresponding to a strain rate of 1 $\times{}$ 10\textsuperscript{-5} s\textsuperscript{-1} for the original length of the sample, were applied during all the experiments in this study. Seal friction along the loading piston was estimated using the axial force before the hit point (Figure \ref{fig:data_processing}a). The strength of the jacket was given by the flow law of \citet{Frost_1982}. The stress on the sample was calculated by the corrected force divided by the cross-sectional area of the sample, which was assumed to increase linearly with deformation with no volumetric expansion. The externally measured displacement data were corrected for elastic distortion of the Inconel piston assembly.\par
    
    Sets of experiments were conducted at three different temperatures under a confining pressure of 400 MPa; one each at room temperature, 200\degree{}C and 350\degree{}C. In each set of experiments, loading was ceased at axial strains of about 0.5\%, 1.0\%, 2.0\%, 4.0\% or 7.5\%. In total, fifteen samples were tested (Table \ref{tab:mech_test_sum}). Run 89 to 7.5\% strain at 200\degree{}C was a previous experiment with the same apparatus at the same strain rate and pressure by \citet{Harbord_2023}. \par

    The hardening modulus ($H$) was computed from the slope ($k$) of the stress-strain curve and the Young's modulus of intact calcite aggregate ($E_{\mathrm{0}}$) by 
    \begin{equation}
        H = (k^{-1} - E_{\mathrm{0}}^{-1})^{-1}.
    \label{eqn: hardening_modulus}
    \end{equation}
    The slope of the stress-strain curve was calculated by linear regression over the last 100\textendash{}200 stress-strain data for each experiment, covering a range of 0.1\textendash{}0.2\% axial strain at about 0.5\%, 1.0\%, 2.0\%, 4.0\% and 7.5\% axial strain before unloading. The Young's modulus $E_0$ was approximated by the Voight-Reuss-Hill average of pure calcite. The Voight and Reuss bounds for calcite's trigonal symmetry were calculated following \citet{Watt_1980}. The elastic moduli of calcite at a confining pressure of 400 MPa and temperatures from 20 to 350\textdegree{}C were extracted from \citet{Dandekar_1968} and \citet{Lin_2013}. The accuracy of $H$ is limited by the measurement of axial load external to the pressure vessel and seals. During the loading process, the steel piston, transmitting force from the actuator to the specimen assembly, horizontally expands due to the Poisson effect, which could lead to an increase in friction at the piston seal. The magnitude of this increase is expected to be minor relative to our correction for seal friction, but might not be negligible in the calculation of hardening modulus. Accordingly, the reported hardening moduli are upper bounds on the true hardening moduli. \par 
    
\subsection{Microstructural characterisation}
    After deformation experiments, the samples were cut into halves from the centre of their top surface along the long axis of each sample. The halved samples were impregnated with epoxy in a vacuum chamber to minimise the possibility of introducing fractures during subsequent polishing. Each epoxy block with an exposed axial plane was ground with \#1000 sandpaper then polished with a Buehler\textsuperscript{TM} MiniMet\textsuperscript{\textregistered} 1000 with progressively finer diamond pastes from 15 \textmu{}m down to 0.25 \textmu{}m. Colloidal silica was used for final polishing. After polishing, the samples were coated with approximately 5 nm of carbon.\par
    
    Microstructural data were acquired on a Zeiss Gemini\textsuperscript{TM} 300 field emission gun scanning electron microscope in the Department of Materials Science and Metallurgy, University of Cambridge. The polished surface was tilted to 70\degree{} within the vacuum chamber. Forescattered electron (FSE) images were acquired at an accelerating voltage of 30 kV, with an aperture 120 \textmu{}m in diameter, at a working distance of 15 mm. The vertical and horizontal position of a scintillator detector with FSE diodes remained the same for all FSE images. EBSD mapping was conducted under the same conditions with a step size of 1.5 \textmu{}m using an Oxford Instruments Symmetry EBSD detector and AZtec\textsuperscript{TM} 4 acquisition software. For each sample, an FSE image with a width of 1.5 mm and height of 1.0 mm was collected near the centre of the polished surface, then an EBSD map of the same size was collected at the same location. \par      
    
    EBSD maps were processed with the MATLAB toolbox MTEX v5.8.1. The crystal symmetry was defined as -3m1 with $a$ = 5 Å, $b$ = 5 Å and $c$ = 17 Å. The reference frame was chosen to be $x$ $||$ $a$*, $y$ $||$ $b$, $z$ $||$ $c$* (Figure~\ref{fig:microstructural_data_processing}a). Here axes $x$, $y$, $z$ form an orthogonal crystal coordinate system as the reference frame. $a$* and $c$* refer to the reciprocal crystal coordinate system. In the coordinate frame of the physical specimen, we denote $Y$ the axis pointing downwards in the direction of the macroscopic axial stress, and $Z$ the axis pointing into the plane of the EBSD maps. Any indexed pixels with fewer than 2 adjacent indexed pixels were removed as noise. Any pixels that were not indexed but had at least one indexed neighbour were filled by with the average orientation of the neighbouring indexed pixels. After data cleaning, calcite grains were reconstructed from EBSD maps with a threshold misorientation angle of 10\textdegree{} between neighbouring indexed pixels. The procedures to acquire key microstructural information, on mechanical twins, lattice curvature, and intragranular microfractures, are described below. \par
    
\subsubsection{Mechanical twins} 
    \label{subsubsec: method, mechanical twins}
\begin{figure}[H]
        \centering
        
        \includegraphics[width=0.7\textwidth]{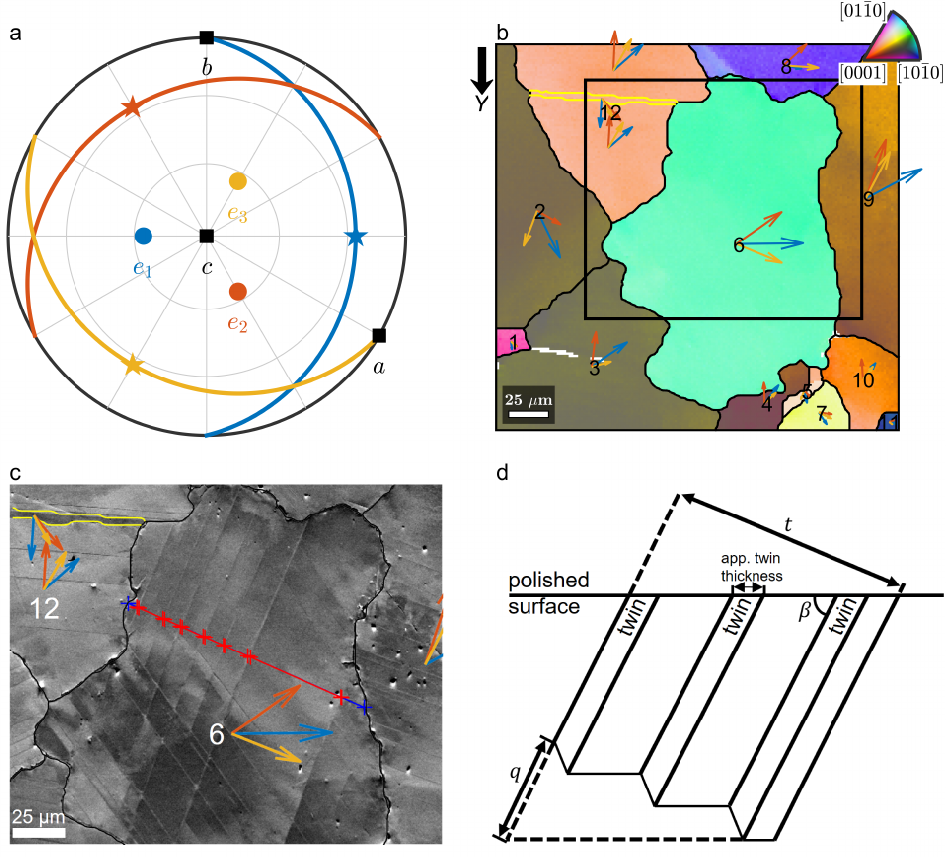}
        \caption{(a) Crystallography of $e$ twins in calcite. Axes $a$, $b$ and $c$ define a fundamental region for calcite. $e$ planes and poles for three twin sets ($e$\textsubscript{1}, $e$\textsubscript{2}, $e$\textsubscript{3}) are coloured as solid lines and circles. The twin shear directions are indicated by stars. For each twin, the $e$ pole, $c$ axis, and corresponding shear direction are co-planar. (b) Selected subarea of the EBSD map of the sample from Run 225. Yellow lines mark the twin boundaries identified by EBSD. Black lines mark grain boundaries. The colour of each pixel is based on the crystal direction parallel to the $Y$ axis (direction indicated by the black arrow) according to the colour key in the top-right corner. This colour scheme is also applied to other orientation maps in this study. Each grain is identified by a number. The three arrows in each grain (or twinned portion) are projections of the three $e$ poles from 3D into the plane of the map. The colour of each arrow corresponds to the colour of $e$ poles in (a). (c) FSE image of part of the mapped area marked by the black rectangle in (b). The apparent twin thicknesses and twin spacings are determined from the coordinates of the red crosses. The distance between the blue crosses records the grain width. The twin boundaries of the measured twin set in Grain 6 are perpendicular to the yellow arrow corresponding to $e$\textsubscript{3}, whereas the other twin set in this grain is perpendicular to the orange arrow for $e$\textsubscript{2}. (d) Schematic illustration of twinned calcite in cross section. $q$ is the shearing distance from twinning, $t$ is the thickness of the grain, and $\beta$ is the inclination angle between the polished surface and a twin-boundary $e$ plane.}
        \label{fig:microstructural_data_processing}
    \end{figure}

     In the crystal coordinates of the hexagonal structural cell for calcite, $e$ twins are formed on \{10$\bar{1}$8\} $e$ planes and shear in $\langle$40$\bar{4}$1$\rangle$ directions (Figure~\ref{fig:microstructural_data_processing}a). We collected data on twin density (i.e., number of twins per unit length), twin spacing, twin thickness, and strain accommodated by twins. \par
     
     In EBSD maps, the twin boundaries are defined by boundaries between neighbouring portions of crystal with misorientation angles of 77.9 $\pm$ 4\textdegree{}, i.e., close to the misorientation angle between a parent grain and $e$ twin in calcite, around crystal axes $\langle$$\bar{1}104$$\rangle$. The tolerance of 4\textdegree{} allows for imprecision in the measurements (on the order of 0.1\textdegree{}, the angular resolution for misorientation angles in EBSD) plus any possible intragranular misorientation that may be present between measurement points due to dislocations and fractures. As the widths of most deformation $e$ twins are thinner than 2 \textmu{}m, the pixels within them are commonly separated and therefore removed by noise reduction, whereas the thicker pre-existing $e$ twins are better resolved (see an example in Figure \ref{fig:microstructural_data_processing}bc). Accordingly, the information on thin mechanical twins is best obtained by manual measurements on FSE images (Figure \ref{fig:microstructural_data_processing}c; \citet{Rutter_2022}). First, a twin boundary is manually traced. Second, a straight line perpendicular to the trace, crossing all the twin boundaries of this set, is drawn. Third, the coordinates of the intersections between twin boundaries and the crossing line are manually picked to calculate the apparent widths of the twins (i.e., twin thickness) and the apparent spacings between two adjacent twins (i.e., twin spacing). \par

    The true twin spacings and twin thicknesses can also be calculated from the locations of twin boundaries. A schematic cross section of twins is plotted in Figure \ref{fig:microstructural_data_processing}d. The apparent twin thicknesses and spacings were acquired as described above. However, twin boundaries, as planes, intersect with the polished surface of the sample by an inclination angle $\beta$. Quantification of the true twin spacing and thickness, that is the distances perpendicular to twin boundaries, requires the value of $\beta$ for each twin set. We obtain $\beta$ by extracting combined information from FSE images and EBSD maps following \citet{Rutter_2022}. The average orientation of the $c$ axis in $X$-$Y$-$Z$ specimen coordinates is obtained from the mean orientation of all pixels in the grain. Thus, the three possible $e$ poles can be computed based on the crystallography in Figure \ref{fig:microstructural_data_processing}a projected onto the 2D map as in Figure \ref{fig:microstructural_data_processing}b and overlapped with the FSE image in Figure \ref{fig:microstructural_data_processing}c. The orthogonal relation between the projection of one of the three $e$ poles and the twin boundaries of interest in Figure \ref{fig:microstructural_data_processing}c determines the orientation of the $e$ pole of the observed twin system. The inclination angle $\beta$ for the twin set of interest is equivalent to the angle between this $e$ pole (normal to the twin boundaries) and the $Z$ axis (normal to the polished surface). The true twin spacing and thickness can then be calculated from the apparent spacing and inclination angle $\beta$. \par

    Figure \ref{fig:microstructural_data_processing}d also illustrates that twinning causes simple shear. The engineering shear strain ($\gamma$) that results from twinning is calculated as the ratio between shearing distance ($q$) and the true width of the grain ($t$). Following the crystallography in Figure \ref{fig:microstructural_data_processing}a, the angular relation between the twins and the host grain is constant. As such, $q$ can be calculated from the sum of twin thicknesses and this constant angle, and therefore the engineering shear strain $\gamma$ resulting from twinning in each grain is given by 
    \begin{equation}
        \gamma= \frac{q}{t} =\frac{2}{t}\sum_{k=1}^{n}t_k\tan(\frac{\alpha}{2}), 
    \label{eqn: engineering_twin_strain}
    \end{equation}
    where $q$ is calculated from the sum of twin thickness ($\sum_{k=1}^{n}t_k$, where $k$ indexes each individual twin) and the constant angular change ($\alpha$ = 37.28\textdegree) of a \{10$\bar{1}$4\} plane as a consequence of twinning \citep{Groshong_1972}.\par

    The optimal directions of compressive ($C$) and tensile ($T$) strains were determined based on the crystal symmetry of calcite. The $C$ axis is oriented at 45\degree{} to the glide direction, and the $T$ axis is oriented at 45\degree{} to the $e$ pole, both in a clockwise manner within the plane containing the $e$ plane normal and $g$ in Figure~\ref{fig:microstructural_data_processing}a. The strain tensor in $C$-$T$ coordinates is
    \begin{equation}
    \boldsymbol{\varepsilon}_{\mathrm{CT}} = 
        \begin{bmatrix}
            \gamma/2 & 0 & 0\\
            0 & -\gamma/2 & 0\\
            0 & 0 & 0
        \end{bmatrix},
    \end{equation}
    which can be converted to a strain tensor in $X$-$Y$-$Z$ specimen coordinates ($\boldsymbol{\varepsilon}_\mathrm{t}$) by
    \begin{equation}
        \boldsymbol{\varepsilon}_\mathrm{t} = \gamma\mathbf{G},  
    \label{eqn:Conel_twin_strain}
    \end{equation}
    where the matrix $\mathbf{G}$ contains the cosines between $C$-$T$ axes and $X$-$Y$-$Z$ axes \citep{Groshong_1972}, transferring the engineering shear strain $\gamma$ in crystal coordinates into the strain tensor $\boldsymbol{\varepsilon}\textsubscript{t}$ in specimen coordinates.\par

    Overall, we collected microstructural information on twin lamellae from 28 to 46 grains from the FSE images and EBSD maps of each sample, including twin density, true twin spacing, true twin thickness and strain accommodated by twins $\boldsymbol{\varepsilon}_\mathrm{t}$. The twin density of each sample is calculated as the total number of measured twins over the total length of measured grain widths. The average true twin spacing and thickness are considered as representative values for each sample. The $YY$ element in the strain tensor $\boldsymbol{\varepsilon}_\mathrm{t}$, denoted as $\varepsilon_\mathrm{tYY}$, indicates the shortening (positive value) or lengthening (negative value) in the direction of macroscopic applied stress of individual grains accommodated by $e$ twins. The average value of $\varepsilon_\mathrm{tYY}$ for each sample is calculated as the weighted average of the $\varepsilon_\mathrm{tYY}$ from the measured grains, using grain areas as weighting factors.\par
    
    For grains with more than one twin set, all the twin information was collected for each twin set. The highest twin density among the measured twin sets in each grain was considered the representative values. The true twin spacing and twin thickness are the averages of all the measurements. $\varepsilon_\mathrm{tYY}$ is the sum of all the $\varepsilon_\mathrm{tYY}$ from different twin sets in these grains. \par
    
    The number of these measured grains is only a small portion of the total number in a bulk sample, and the extent of twinning varies among the grains. Accordingly, to evaluate how representative these measurements are of the characteristics of the bulk sample, we use a bootstrap method to calculate the standard deviation of distributions from resampling of these measurements. The measured grains of each sample are randomly assigned into a test set. The twin information of this test set is calculated as described above. This procedure of randomly choosing grains and calculating twin information is run iteratively 10,000 times for each sample. This resampling strategy provides a normal distribution of each type of twin information. The standard deviation of this distribution is considered as the uncertainty of these measurements for the bulk sample.\par

\subsubsection{Lattice curvature}
    Our approach to evaluate lattice curvature associated with dislocations is introduced in this subsection. EBSD measurements provide the crystal orientation at each point in specimen coordinates. The misorientation angle $\mathrm{d}\theta$ between any two points is defined as a rotation angle about a common axis determined as a unit vector [$uvw$] in crystal coordinates \citep[c.f., Figure 4 in][]{Muransky_2019}. Lattice curvature $\kappa_{ij}$ is defined as the infinitesimal misorientation $\mathrm{d}\theta_{i}$ across an infinitesimal displacement $\mathrm{d}u_{j}$ along an axis in specimen coordinates ($i$ = 1, 2, 3 for $x$, $y$, $z$ axes, respectively; $j$ = 1, 2, 3 for $X$, $Y$, $Z$ axes, respectively) as
    \begin{equation}
        \kappa_{ij} = \frac{\mathrm{d}\theta_{i}}{\mathrm{d}u_{j}}, 
    \end{equation}
    where 
    \begin{equation}
        \mathrm{d}\theta_{1} = |\mathrm{d}\theta|\frac{u}{\sqrt{u^2+v^2+w^2}},
        \mathrm{d}\theta_{2} = |\mathrm{d}\theta|\frac{v}{\sqrt{u^2+v^2+w^2}},
        \mathrm{d}\theta_{3} = |\mathrm{d}\theta|\frac{w}{\sqrt{u^2+v^2+w^2}}. 
    \end{equation}
    With the assumption that spatial gradients in elastic strain are negligible, the lattice curvature $\kappa_{ij}$ results from the presence of dislocations \citep{Nye_1953}. Accordingly, in theory, it is possible to acquire the density of each type of dislocation from lattice curvature. \par
    
    In practice, EBSD maps can only offer limited information on lattice curvature and thus dislocations. Firstly, EBSD maps are collected with a finite step size (here, 1.5 \textmu{}m). As such, dislocations with opposite signs of Burgers vector between two neighbouring pixels (i.e., statistically stored dislocations, SSD) do not contribute to the lattice curvature from EBSD maps. Instead, only the portion of dislocations with a nonzero sum of Burgers vectors can be captured, termed geometrically necessary dislocations (GND). Secondly, curvature in the direction of $Z$ axis ($j$ = 3; normal to the map) cannot be accessed with a 2D map, and thus only six elements of $\kappa_{ij}$ in the 2D plane containing the $X$ and $Y$ axes can be directly measured by EBSD. Thirdly, calcite as high-symmetry crystal contains more than six slip systems \citep{de_1991}, thus the solution of dislocation densities for individual dislocation types from the incomplete lattice curvature is non-unique. Lastly and most importantly, the angular resolution of the conventional EBSD data (at the order of 0.1\textdegree{}) can obscure GND in pixel-level calculations, in which the length scale over which curvature is measured is typically relatively short. Accordingly, instead of resolving dislocation density from pixel-to-pixel measurements, we used a simplified relation between misorientation at the grain scale and GND density from \citet{Ashby_1970} to estimate $\rho_{\mathrm{GND}}$,  
    \begin{equation}
        \rho_{\mathrm{GND}}= \frac{\theta_\mathrm{m}}{bd},
    \label{eqn: GND_density_Ashby}
    \end{equation}
    where $\theta_\mathrm{m}$ is the misorientation angle at the grain scale, $b$ is the average length of the available Burgers vectors, and $d$ is grain size. We used the grain orientation spread (GOS) as an estimate of $\theta_\mathrm{m}$, calculated by 
    \begin{equation}
        \theta_\mathrm{m} = \mathrm{GOS} = \frac{1}{n}\sum_{k=1}^{n}\theta_{k},
    \label{eqn: GOS}
    \end{equation}
    where $n$ is the number of pixels in a grain, and $\theta_{k}$ is the misorientation angle between the orientation of the $k$th pixel and the mean orientation of the grain. To be consistent with using GOS as the measure of misorientation at the grain scale, $d$ in Equation. (\ref{eqn: GND_density_Ashby}) is also replaced by the average of the distance between each pixel and the grain centroid, $d_\mathrm{c}$. Overall, we have 
    \begin{equation}
        \rho_\mathrm{GND} = \frac{\mathrm{GOS}}{bd_\mathrm{c}}
    \label{eqn: GND_density_this_study}
    \end{equation}
    as an estimate of the minimum density of GND for each grain. As such, by increasing the length scale over which curvature is measured up to the scale of the grain size, we can improve the signal-to-noise ratio for estimates of lattice curvature and GND density. To acquire $\rho_\mathrm{GND}$ for the entire EBSD map as the representative $\rho_\mathrm{GND}$ for the sample, $\rho_\mathrm{GND}$ of each grain is multiplied by the individual grain area, the sum of which is normalised by the whole map area. Each map contains 200\textendash{}300 grains, and no manual operation or non-uniqueness is involved in data acquisition. \par

\subsubsection{Intragranular microfractures}
    Fractures visible in FSE images were traced in ImageJ by recording the coordinates of the ends of a fracture and the nodes where the fracture direction changes. The tracing results were analysed with the MATLAB\textsuperscript{\textregistered} toolbox FracPaQ v2.8 \citep{Healy_2017}. As most grain boundaries are open in Carrara marble after decompression and polishing \citep{Harbord_2023}, only intragranular fractures of the experimentally deformed samples were traced. Fracture intensity is calculated as the sum of fracture lengths over the image area. \par

    This approach of tracing fractures is highly dependent on the image quality, which is influenced by many factors, including the thickness of the carbon coat and the focus of the electron beam. Also, this traced map represents only a portion of the sample. To evaluate whether the fracture intensity determined from the image area is representative of the bulk sample, we divided each FSE image into 100 strips. The width of each strip is the same as the width of the complete image (1.5 mm), and the height of each strip is 1/100 of the image height (i.e., 10 \textmu{}m). The traced fractures were assigned into each strip based on their node locations. This subdivision revealed that with 25 randomly chosen strips the fracture intensity is close (within about $\pm$ 5\%) to the fracture intensity of the whole map. For this reason, 100 testing blocks, each with a quarter of the image area, were randomly distributed onto each FSE image. The distribution of fracture intensity from all the testing blocks is used for calculating a standard deviation as the uncertainty of the measurement of fracture intensity for each sample. \par

    Cleavage fractures were identified by an approach similar to that used to identify twin sets. The normal of each of the three cleavage planes (i.e., $r$ planes, \{10$\bar{1}$4\}) were projected onto each grain in the 2D EBSD maps. When the section between two nodes of a traced fracture is at an angle of 90 $\pm$ 10\textdegree{} to one of the projected cleavage normals (where the range allows for imprecision in tracing), the section of traced fracture is assigned as cleavage. \par

\section{Results}
\label{sec:results}
\subsection{Mechanical testing}
\begin{figure}[H]
        \centering
        \includegraphics[width=0.8\textwidth]{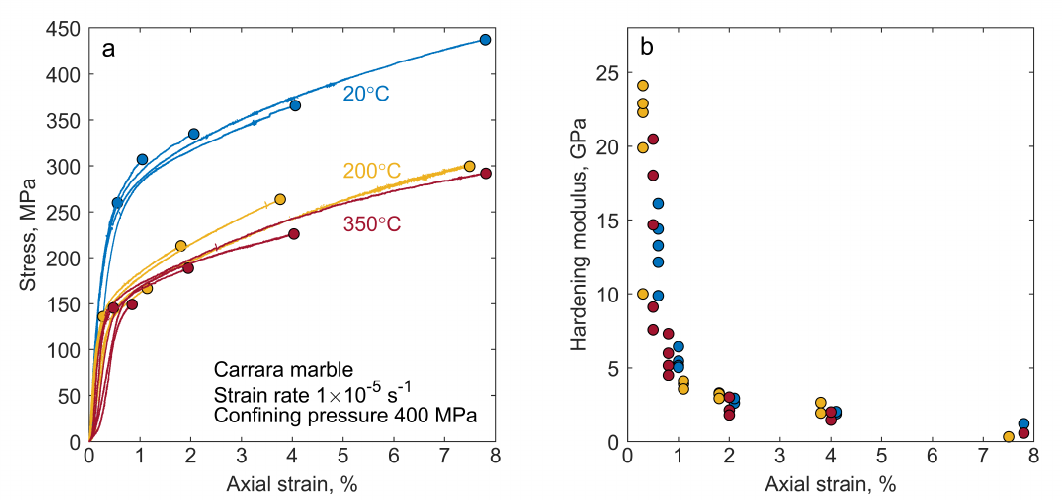}
        \caption{(a) Stress-strain data at three experimental temperatures. Temperatures are colour coded. The end of loading in each experiment is highlighted by a circle, summarised as $\varepsilon$ and $\sigma$ in Table \ref{tab:mech_test_sum}. (b) Hardening modulus derived from each stress-strain curve in (a) and Young's modulus from Voigt-Reuss-Hill average stiffness of calcite at the experimental conditions. The colours in (b) follow the colour code in (a).}
        \label{fig:strain_series_data}
    \end{figure}

The mechanical testing results are summarised in Table \ref{tab:mech_test_sum} and presented in Figure \ref{fig:strain_series_data}. Temperature has an overall weakening effect on the samples. With the difference in temperature between 20\textdegree{}C and 200\textdegree{}C, the stress at a given axial strain decreases by about 100\textendash{}150 MPa. However, the difference in temperature between 200\textdegree{}C and 350\textdegree{}C produces only a modest decrease in strength. \par

The mechanical behaviour of Carrara marble exhibits systematic strain hardening with common features observed at the three temperatures. The stress-strain data at the three temperatures display elastic segments at strains \textless{} 0.2\%{}. At about 0.5\textendash{}1\% axial strain, the slope of the stress-strain curves decreases markedly, and beyond about 1\% axial strain the slope continues to decrease but at a lesser rate. These trends are quantified by the hardening modulus, which exhibits a sharp decline with increasing axial strain up to 1\textendash{}2\%, followed by a more gradual decrease between 2\% and 8\% axial strain (Figure \ref{fig:strain_series_data}b). The hardening modulus does not exhibit a resolvable temperature dependence.\par

    \begin{table}[H]
        \small
        \centering
        \begin{threeparttable}
        \caption{Summary of mechanical testing}
        \centering
        \label{tab:mech_test_sum}
        \begin{tabular}{ccccc}
            \hline 
            Run & Temperature, \degree{}C & $\varepsilon$\tnote{1}, \% & $\sigma$\tnote{1}, MPa & Hardening modulus\tnote{2}, GPa\\
            \hline
            241 & 20 & 0.6 & 260 & 15.1 \\
            239 & 20 & 1.0 & 307 & 8.5  \\
            238 & 20 & 2.1 & 335 & 2.4  \\
            237 & 20 & 4.1 & 366 & 1.9  \\
            242 & 20 & 7.8 & 437 & 1.2  \\

            222 & 200 & 0.3 & 136 & 26.7 \\
            221 & 200 & 1.1 & 167 & 4.3  \\
            225 & 200 & 1.8 & 213 & 3.2  \\
            223 & 200 & 3.8 & 264 & 2.9  \\
            89  & 200 & 7.5 & 300 & 0.8  \\

            249 & 350 & 0.5 & 146 & 10.4 \\
            247 & 350 & 0.8 & 149 & 4.7  \\
            246 & 350 & 2.0 & 189 & 2.7  \\
            248 & 350 & 4.0 & 226 & 1.3  \\
            244 & 350 & 7.8 & 292 & 0.6  \\
            \hline
        \multicolumn{5}{l}
        
       \end{tabular} 
        
        \begin{tablenotes}
            \item [1] {\footnotesize $\varepsilon$ and $\sigma$ are strain and stress data at the end of loading, respectively.}
            \item [2] {\footnotesize Hardening modulus here was calculated from the final 0.1\% to 0.2\% strain before unloading by eqn. (\ref{eqn: hardening_modulus}).}            
        \end{tablenotes}
        \end{threeparttable}
    \end{table}

\subsection{Microstructures}
\begin{figure}[H]
        \centering
        \includegraphics[width=0.8\textwidth]{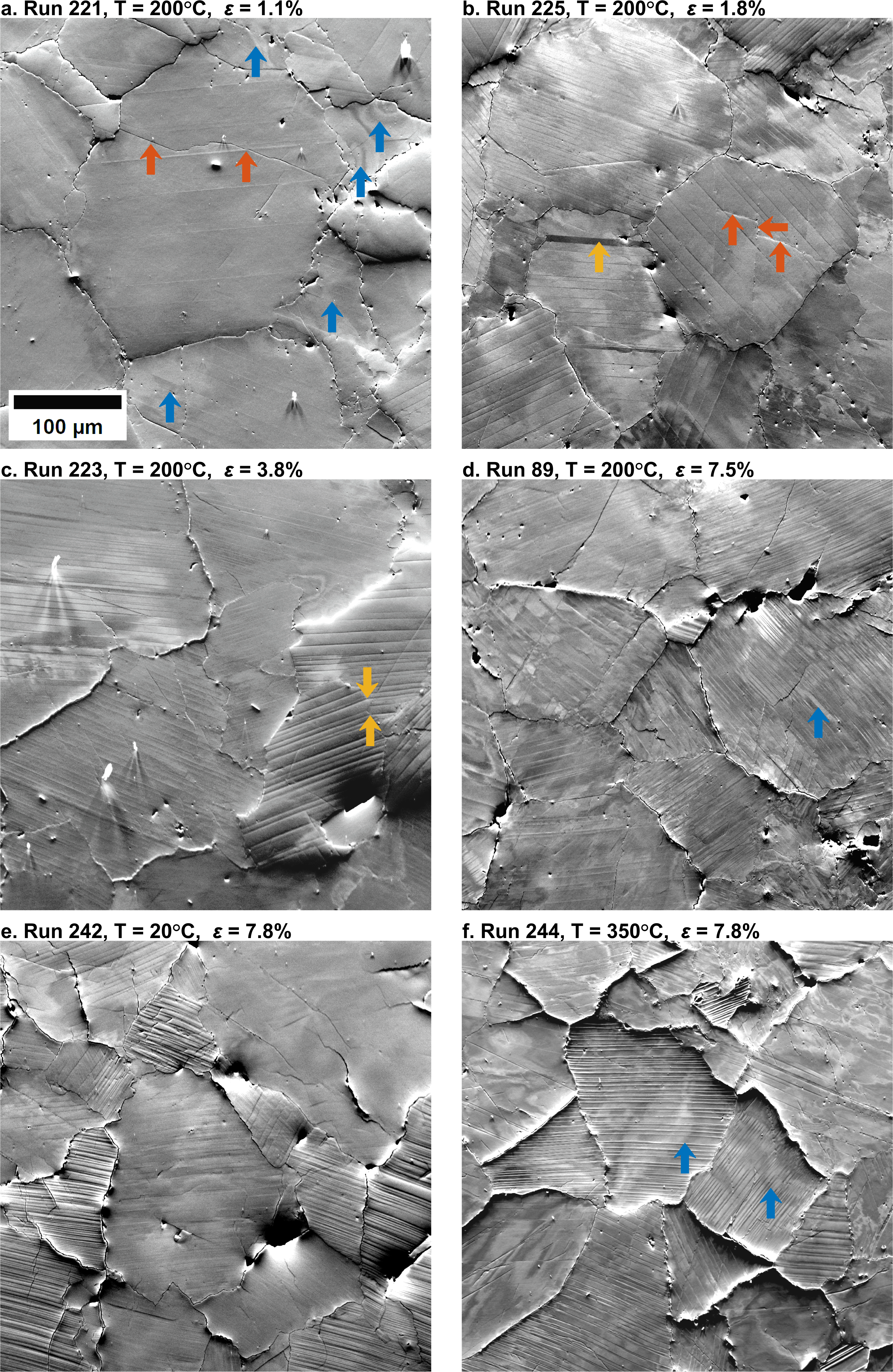}
        \caption{Forescattered electron (FSE) images of the tested marble samples. The scale bar in (a) is common to all the subfigures. (a) The sample deformed by 1.1\% at 200\textdegree{}C from Run 221. Lattice curvature is evident as variation in grey scale, with examples indicated by blue arrows. The straight fractures indicated by orange arrows are possibly cleavage fractures. (b) The sample deformed by 1.8\% at 200\textdegree{}C from Run 225. Fractures possibly induced by twinning are indicated by orange arrows. A pre-existing twin is indicated by the yellow arrow. (c) The sample deformed by 3.8\% at 200\textdegree{}C from Run 223. The yellow arrows indicate twins induced by twins in neighbouring grains. (d) Sample deformed by 7.5\% at 200\textdegree{}C from Run 89. The blue arrow points to lattice curvature segmented by twin boundaries. (e) Sample deformed by 7.8\% at room temperature from Run 242. (f) Sample deformed by 7.8\% at 350\textdegree{}C from Run 244. The blue arrow points to lattice curvature across twin boundaries.}
        \label{fig:marble_image}
    \end{figure}

Inelastic strain in the deformed marble samples is recorded as mechanical twins, lattice curvature, and microfractures. As strain increased, these three types of microstructures developed as demonstrated by Figures \ref{fig:marble_image}a\textendash{}\ref{fig:marble_image}d for marble samples tested at a temperature of 200\textdegree{}C. Fractures are mostly intragranular. There is no sign of lattice curvature induced by cracks, nor vice versa. Lattice curvature was found to be strong at grain contacts at the early stage of deformation, indicated by blue arrows in Figure \ref{fig:marble_image}a. As strain develops, lattice curvature is either segmented by twin boundaries (indicated by the blue arrows in Figure \ref{fig:marble_image}d) or continues uninterrupted across twin boundaries (indicated by the blue arrows in Figure \ref{fig:marble_image}f). The FSE images do not exhibit substantial differences in microstructures among samples deformed at temperatures of 20\textdegree{}C, 200\textdegree{}C or 350\textdegree{}C to the same strain (e.g., a strain of 7.5\% in Figures \ref{fig:marble_image}d\textendash{}\ref{fig:marble_image}f). From the mean orientations of grains in EBSD maps, the $M$ indices of crystallographic preferred orientations, ranging from 0.003 to 0.014 (Table \ref{tab:microstructural_summary}), are independent of macroscopic strain and experimental temperature. The low $M$ indices indicate a lack of crystallographic preferred orientation.\par

    \begin{landscape}
        \begin{table}[htb]
            \tiny
            \begin{threeparttable}
                \caption{Summary of microstructural information}
                \centering
                \label{tab:microstructural_summary}
                \begin{tabular}{cccccccccc}
            
            \hline 
            Run 
            & Temperature, \degree{}C 
            & $\varepsilon$\textsubscript, \% 
            & \makecell{Grains measured  \\ for twin information}  
            & \makecell{Twin density \\ ($\pm$ uncertainty), mm\textsuperscript{-1}}
            & \makecell{True Twin spacing \\ ($\pm$ uncertainty), \textmu{}m}
            & \makecell{$\varepsilon_\mathrm{tYY}$ \\ ($\pm$ uncertainty), \% }
            & \makecell{GND density, \\ 1$\times$10\textsuperscript{12} m\textsuperscript{-2}}
            & \makecell{Fracture intensity, \\ ($\pm$ uncertainty) mm\textsuperscript{-1}}
            & $M$ index \tnote{1} \\
            
            \hline
            241 & 20  & 0.6 & 31 & 50.5 (5.8)   & 11.6 (1.1) & 0.4 (0.3) & 0.54 & 1.7 (0.5)  & 0.009 \\
            239 & 20  & 1.0 & 28 & 59.1 (6.4)   & 7.9  (0.9) & 1.3 (0.2) & 0.57 & 3.0 (0.6)  & 0.008 \\
            238 & 20  & 2.1 & 30 & 126.3 (11.8) & 4.9  (0.3) & 1.9 (0.3) & 0.69 & 7.9 (1.0)  & 0.008 \\
            237 & 20  & 4.1 & 46 & 264.6 (17.2) & 2.5  (0.2) & 3.5 (0.5) & 0.91 & 14.9 (2.2) & 0.006 \\
            242 & 20  & 7.8 & 41 & 276.8 (20.4) & 2.1  (0.2) & 4.0 (0.9) & 1.29 & 17.5 (2.8) & 0.014 \\
            \hline
            222 & 200 & 0.3 & 34 & 20.0 (1.9)   & 30.0 (2.6) & 0.4 (0.1) & 0.53 & 1.0 (0.6) & 0.014 \\
            221 & 200 & 1.1 & 36 & 62.3 (4.7)   & 10.7 (0.8) & 1.0 (0.2) & 0.55 & 2.7 (1.0) & 0.007 \\
            225 & 200 & 1.8 & 31 & 109.9 (11.8) & 6.4  (0.6) & 1.5 (0.3) & 0.56 & 5.6 (0.9) & 0.011 \\
            223 & 200 & 3.8 & 43 & 153.8 (10.8) & 4.4  (0.3) & 2.0 (0.5) & 0.60 & 6.8 (0.7) & 0.011 \\
            89  & 200 & 7.5 & 41 & 200.8 (16.5) & 3.3  (0.3) & 3.8 (0.7) & 1.64 & 11.6(1.3) & 0.006 \\
            \hline
            249 & 350 & 0.5 & 30 & 33.4 (2.5)   & 18.4 (1.7) & 0.3 (0.1) & 0.63 & 2.0 (0.5) & 0.006 \\
            247 & 350 & 0.8 & 29 & 31.0 (4.2)   & 17.5 (2.3) & 0.4 (0.1) & 0.55 & 3.0 (0.5) & 0.010 \\
            246 & 350 & 2.0 & 30 & 106.0 (8.3)  & 6.5  (0.5) & 1.9 (0.4) & 0.68 & 5.8 (1.8) & 0.014 \\
            248 & 350 & 4.0 & 45 & 138.0 (10.6) & 4.7  (0.3) & 2.6 (0.6) & 0.72 & 8.0 (1.3) & 0.011 \\
            244 & 350 & 7.8 & 44 & 226.0 (15.1) & 2.8  (0.2) & 4.3 (0.6) & 1.59 & 9.8 (1.7) & 0.003 \\
            \hline
        %\multicolumn{5}{l}
            \end{tabular} 
        
            \begin{tablenotes}
                \tiny
                \item [1] {\tiny $M$ index for the strength of crystallographic preferred orientation of all the grains in each EBSD map \citep{Skemer_2005}.}
            \end{tablenotes}
            \end{threeparttable}
        \end{table}
    \end{landscape}
    
More qualitative descriptions of the microstructures of Carrara marble undergoing semi-brittle deformation can be found in previous studies \citep[e.g.,][]{Olsson_1976, Fredrich_1989, Rybacki_2021, Harbord_2023}. In the following, we focus on quantitative characterisation of mechanical twins, lattice curvature, and intragranular microfractures. \par

\begin{figure}[H]
        \centering
        \includegraphics[width=0.8\textwidth]{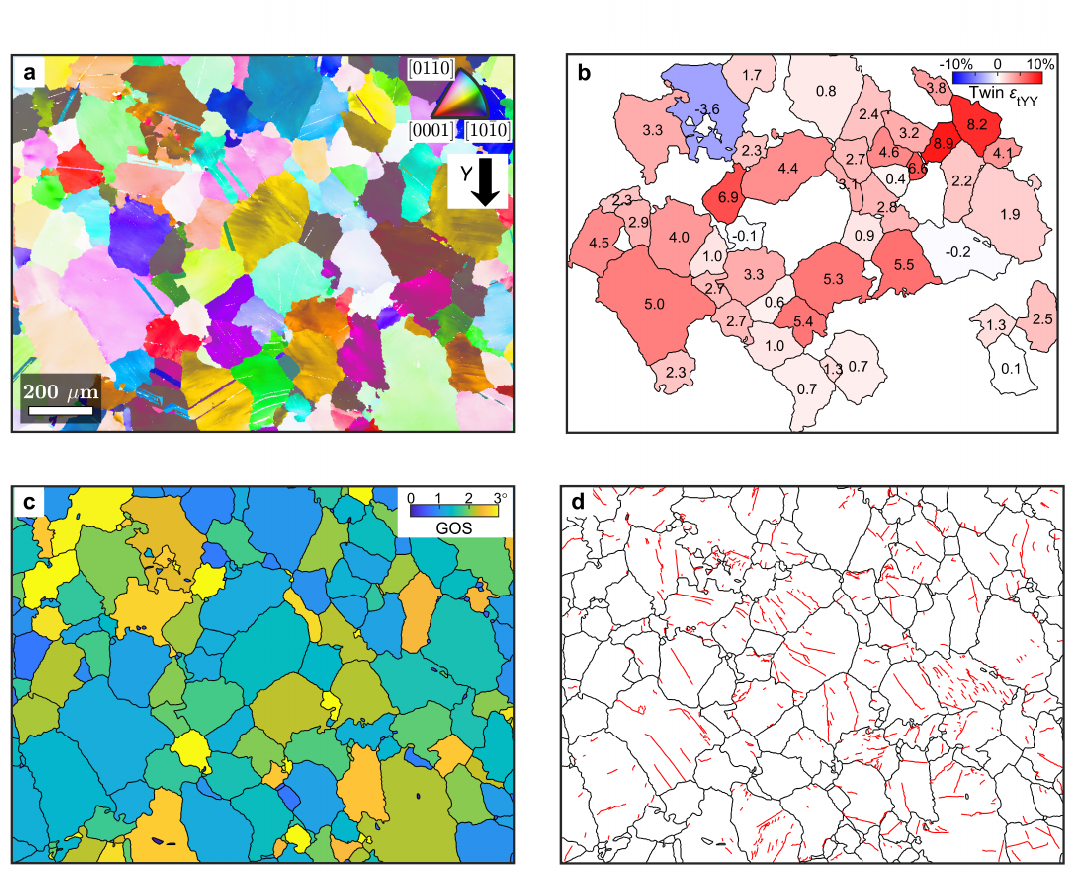}
        \caption{Example of microstructure maps for the sample deformed to a strain of 4.0\% at a temperature of 350\textdegree{}C from Run 248. (a) Orientation map from EBSD coloured by the inverse pole figure for the $Y$ direction (i.e., parallel to the shortening direction indicated by the black arrow.) (b) Map of twin strain of selected grains. The positive/negative value of strain is marked on each grain and presented by red/blue colours. (c) Map of GOS of each grain. (d) Traced intragranular microfractures in red. Reconstructed grain boundaries are in black. The scale bar in subplot (a) is shared with the other three subplots. }
        \label{fig:maps}
    \end{figure}

\begin{figure}[H]
        \centering
        \includegraphics[width=1\textwidth]{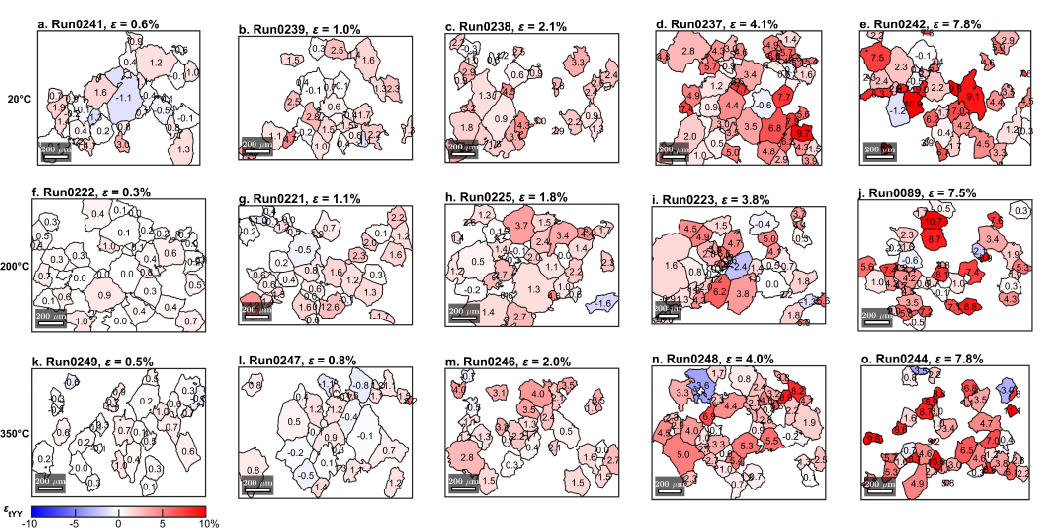}
        \caption{Maps for twin strain $\varepsilon_\mathrm{tYY}$ of selected grains from all the samples deformed at temperatures of 20\textdegree{}C (a–e), 200\textdegree{}C (f–j) or 350\textdegree{}C (k–o). Positive (negative) values of percentage strain for compression (extension) are marked on each grain and presented as red (blue) colours. From left to right, the macroscopic strain $\varepsilon$ increases from 0.5\% to 7.5\%. All the subplots share the same colour bar.}
        \label{fig:maps_all_twin}
    \end{figure}

\begin{figure}[H]
        \centering
        \includegraphics[width=1\textwidth]{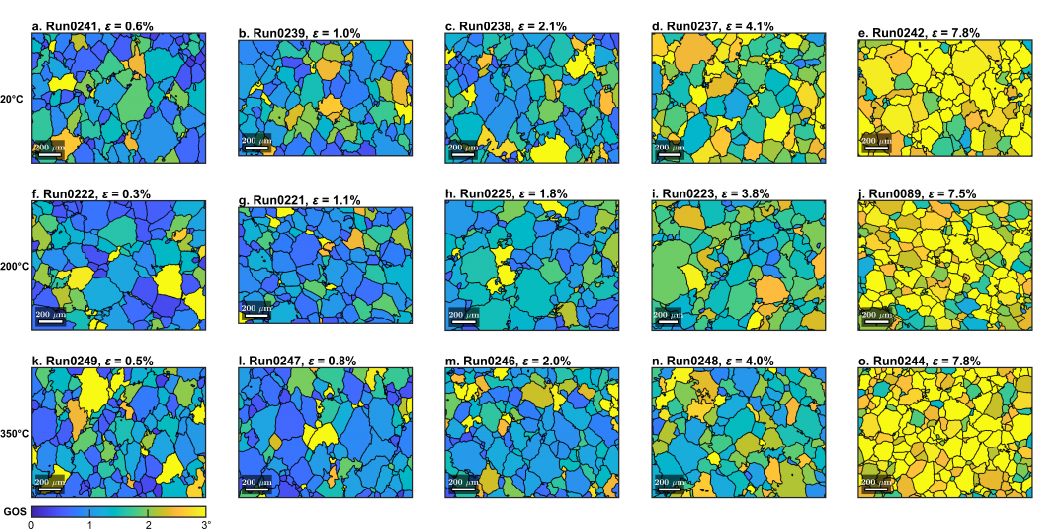}
        \caption{Maps of grain orientation spread (GOS) from all the samples deformed at temperatures of 20\textdegree{}C (a-e), 200\textdegree{}C (f-j) or 350\textdegree{}C (k-o). All the subplots share the same colour bar.}
        \label{fig:maps_all_GOS}
    \end{figure}

\begin{figure}[H]
        \centering
        \includegraphics[width=1\textwidth]{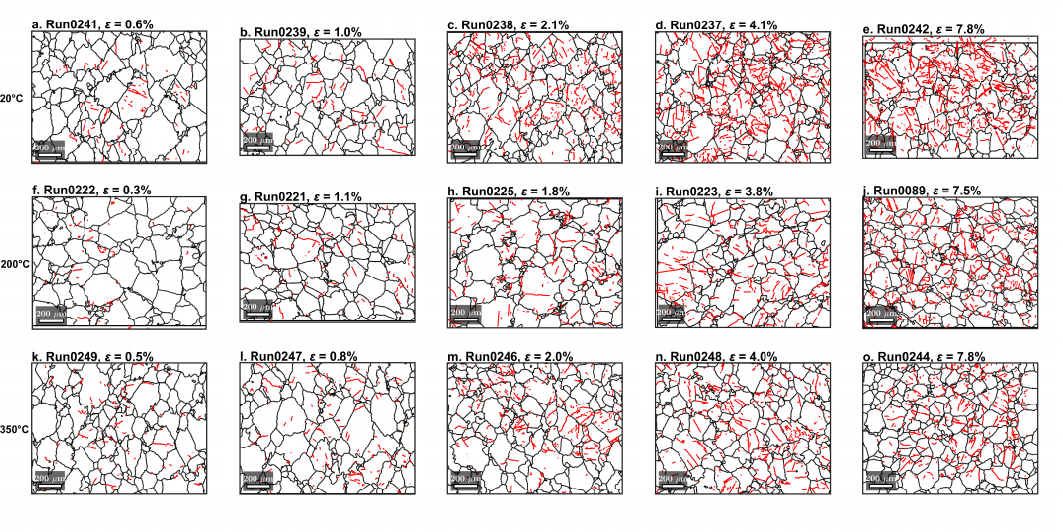}
        \caption{Traced intragrnular microfractures (in red) from all the samples deformed at temperatures of 20\textdegree{}C (a-e), 200\textdegree{}C (f-j) or 350\textdegree{}C (k-o). The size of each FSE image for tracing may be modestly mismatched with the size of the corresponding EBSD map (e.g., subplots e, f, i, o). The offset sections were not considered further in the analysis.}
        \label{fig:maps_all_frac}
    \end{figure}

An example of quantitative characterisation of the sample from Run 248 is presented in Figure \ref{fig:maps}. This sample was deformed to a strain of 4.0\% at a temperature of 350\textdegree{}C. The variation of lattice orientation is plotted in the orientation map in Figure \ref{fig:maps}a. The axial strain accommodated by twins, presented in Figure \ref{fig:maps}b, is variable among grains, with maximum axial strains ranging from -3.6\% (i.e., net extension in the direction of the maximum applied load) to +8.9\%. Similarly, the GOS map and traced fractures (Figures \ref{fig:maps}c and \ref{fig:maps}d, respectively) also illustrate heterogeneous GOS and fracture intensity among grains. The other samples share similar heterogeneity in twin strain, GOS and microcrack distribution. The compiled displays of all the maps indicate increases of twin strain, GOS, and fracture intensity with increasing macroscopic axial strain (Figures \ref{fig:maps_all_twin}, \ref{fig:maps_all_GOS}, and \ref{fig:maps_all_frac}). In the following, we quantitatively examine how these microstructures evolve with increasing macroscopic strain at the three tested temperatures.\par

\subsubsection{Mechanical twins}
\begin{figure}[H]
        \centering
        \includegraphics[width=0.8\textwidth]{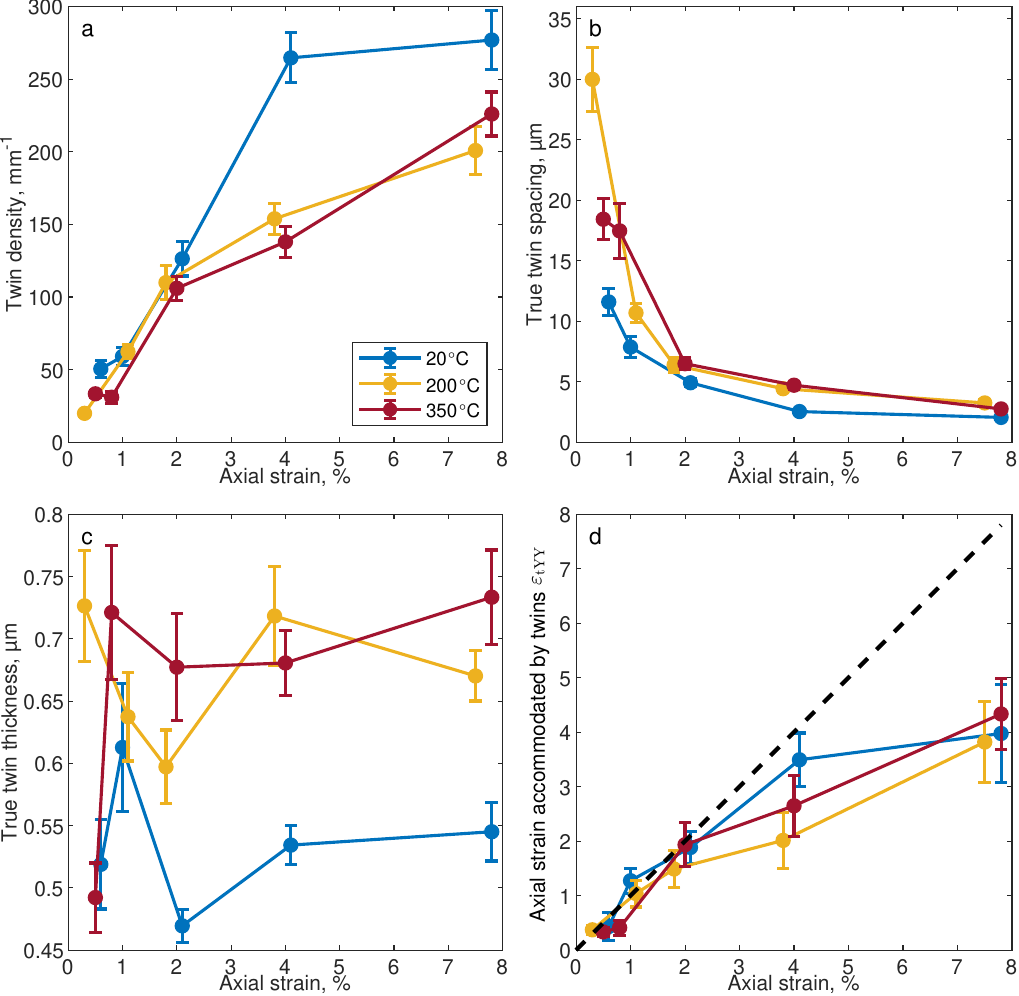}
        \caption{(a) Twin density of each tested sample, calculated as the total number of measured twins over the total length of measured grain widths. (b) Average true twin spacing. The data here and in subplot (c) are corrected by the inclination angle $\beta$ for each set of twins. For data above 4\% axial strain, the error bar covers a range around, or smaller than, the diameter of the markers. (c) Average true twin thickness. (d) Component of the twin-strain tensor parallel to the loading direction. The error bars are uncertainties of measurements introduced in Section \ref{subsubsec: method, mechanical twins}.}
        \label{fig:twin_data}
    \end{figure}

The characteristics of mechanical twins in marble are summarised in Figure \ref{fig:twin_data}. During the first 2\% axial strain, the twin densities of samples deformed at the three different temperatures increase at similarly steep rates of approximately 51 mm$^{-1}$ per 1\% axial strain (Figure \ref{fig:twin_data}a). At room temperature, twin density keeps increasing at a similar rate up to 4\% axial strain, but saturates between 4\% to 8\% axial strain. For samples deformed at temperatures of 200\textdegree{}C and 350\textdegree{}C, the accumulation rate significantly decreases after 2\% axial strain to approximately 21 mm$^{-1}$ per 1\% axial strain, but nonetheless the twin density continues to gradually increase rather than saturating within the strain range of our experiments. The twin densities in samples deformed at a temperature of 200\textdegree{}C are within uncertainty of those in samples deformed at 350\textdegree{}C. \par

The true twin spacing decreases with axial strain at each temperature (Figure \ref{fig:twin_data}b). True twin spacing ranges from approximately 10 \textmu{}m to 30 \textmu{}m at low strain, and decreases to around 5 \textmu{}m at 2\% strain. From 1\% to 4\% strain, the twin spacing is slightly less at room temperature than at 200\textdegree{}C and 350\textdegree{}C. At about 8\% strain, the twin spacing data at the three temperatures are approximately the same at about 2.5 \textmu{}m. The true twin thickness is independent of axial strain, and is generally slightly less at room temperature (around 0.5 \textmu{}m) than that at higher temperatures (around 0.7 \textmu{}m) (Figure \ref{fig:twin_data}c). \par

At axial strains $\le$ 2\%, the average axial strain accommodated by mechanical twins, $\varepsilon{}_\mathrm{tYY}$, is equal to the imposed axial strain across the three temperatures (Figure \ref{fig:twin_data}d). However, between 4\% and 8\% axial strain, $\varepsilon{}_\mathrm{tYY}$ is typically only approximately half of the imposed strain.\par

\subsubsection{Lattice curvature}
\begin{figure}[H]
        \centering
        \includegraphics[width=0.8\textwidth]{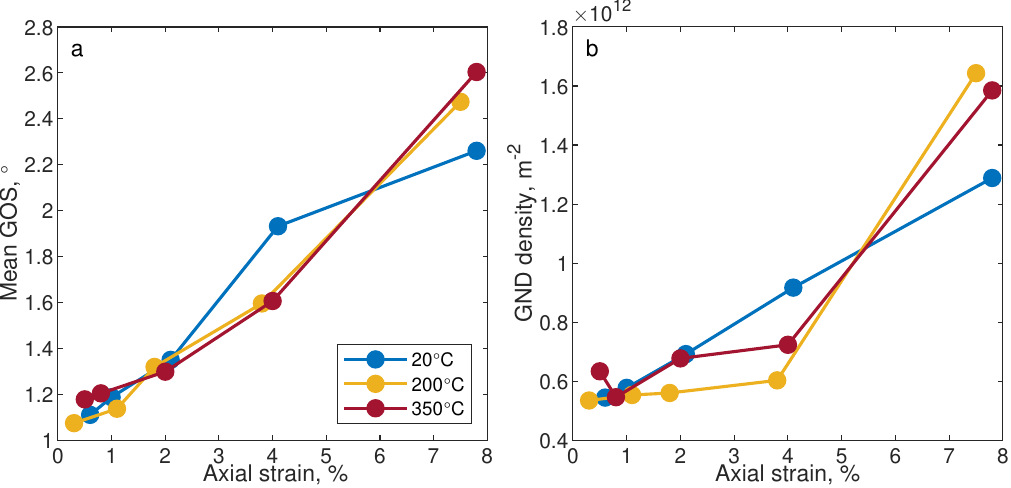}
        \caption{(a) Mean GOS against axial strain. (b) GND density, estimated from GOS and grain size, against axial strain. }
        \label{fig:GOS_GND_data}
    \end{figure}

The mean GOS increases with strain at all the tested temperatures, from around 1\textdegree{} at $<$ 1\% strain to 2.2\textdegree{}\textendash{}2.6\textdegree{} at 7.5\% strain (Figure \ref{fig:GOS_GND_data}a). The rate of increase in GOS increases with strain at 200\textdegree{}C and 350\textdegree{}C, but is approximately linear at room temperature. The corresponding GND densities also increase with increasing strain (Figure \ref{fig:GOS_GND_data}b). At the smallest strains of around 0.5\%{}, the GND densities are similar between the three experimental temperatures at around 0.6 $\times{}$ 10\textsuperscript{12} m\textsuperscript{-2}. At room temperature, the increase is linear up to 1.3 $\times{}$ 10\textsuperscript{12} m\textsuperscript{-2} at 8\% strain. In contrast, at 200\textdegree{}C and 350\textdegree{}C, the increase in GND density is minor up to 4\% strain, reaching only around 0.7 $\times{}$ 10\textsuperscript{12} m\textsuperscript{-2}, but then sharply rises up to 1.6 $\times{}$ 10\textsuperscript{12} m\textsuperscript{-2} at 7.5\% strain.

\subsubsection{Intragranular microfractures}
\begin{figure}[H]
        \centering
        \includegraphics[width=0.8\textwidth]{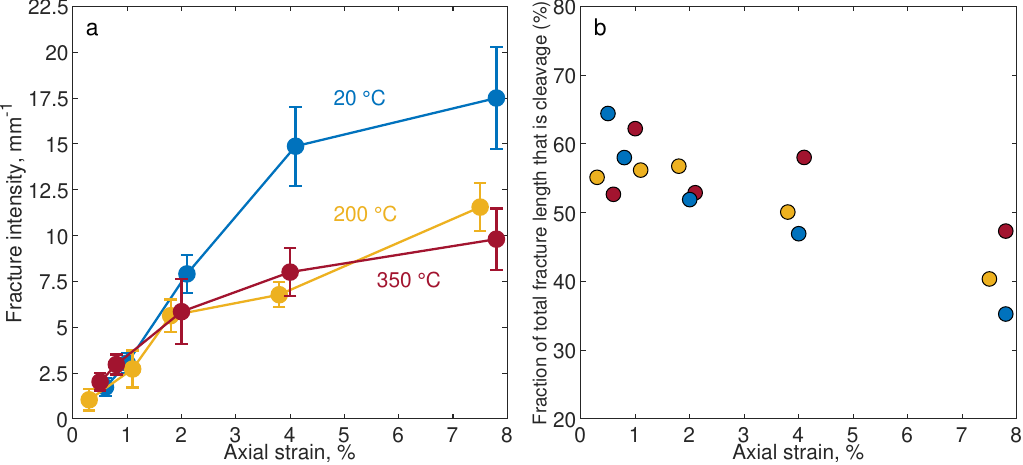}
        \caption{(a) Fracture intensity against axial strain at different temperatures. (b) Fraction of total fracture length that is cleavage against axial strain. The colours of markers represent temperatures, shared in the two subplots. }
        \label{fig:fracture_data}
    \end{figure}

 At axial strains $\le$ 2\%, fracture intensity exhibits linear increases with strain at rates of approximately 3.5 mm$^{-1}$ per 1\% axial strain at all three temperatures in Figure \ref{fig:fracture_data}a. Above 2\% strain, fracture intensities at room temperature are greater than those at the higher temperatures. Specifically, at room temperature, the slope of fracture intensity remains steady between 2\% and 4\% strain but decreases between 4\% and 8\% axial strain. At 200\textdegree{}C and 350\textdegree{}C, fracture intensity increases more gradually from 2\% to 8\% strain than at room temperature, but is similar between these two higher temperatures. \par

 From 0.5\% to 2\% axial strain, more than 50\% of the length of fractures is attributable to cleavage fractures. This portion is independent of temperature. From 4\% to 8\% strain, the proportion of cleavage fractures drops below 50\%. During these late stages of deformation, the proportion decreases more rapidly with strain with decreasing temperature (Figure \ref{fig:fracture_data}b). \par

\section{Discussion}

\subsection{General characteristics of strength evolution}

\begin{figure}[H]
        \centering
        \includegraphics[width=0.9\textwidth]{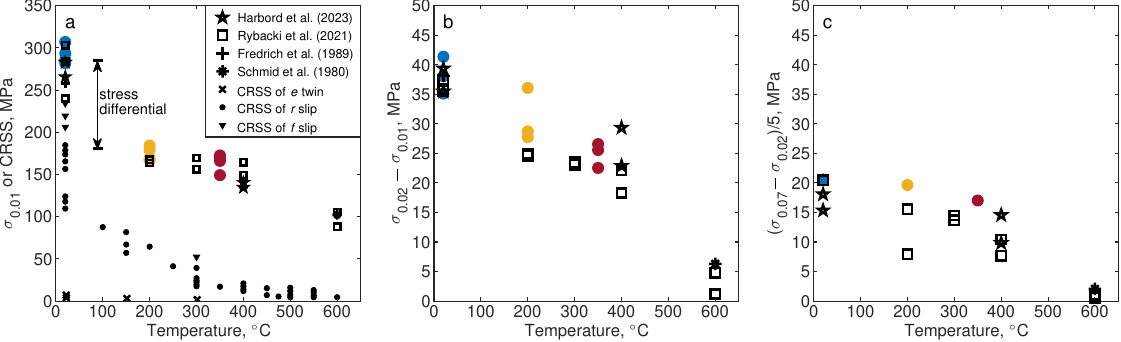}
        \caption{Mechanical data extracted from experiments on Carrara marble at pressures (300\textendash{}400 MPa) and strain rates (1$\times$10\textsuperscript{-6}\textendash{}5$\times$10\textsuperscript{-5} s$^{-1}$) similar to those in our experiments. Coloured markers correspond to those in Figure \ref{fig:strain_series_data}a. $\sigma$\textsubscript{0.01}, $\sigma$\textsubscript{0.02} and $\sigma$\textsubscript{0.07} are the absolute values of macroscopic stress at 1\%, 2\% and 7\% axial strain, respectively. (a) Macroscopic stress at 1\% axial strain, $\sigma$\textsubscript{0.01}, from this and previous studies \citep{Schmid_1980, Fredrich_1989, Rybacki_2021, Harbord_2023} along with the critical resolved shear stresses (CRSS) for $e$ twinning, $r$\{10$\bar{1}$4\}$\langle$$\bar{2}021$$\rangle$\textsuperscript{$\pm$} slip, and $f$\{$\bar{1}$012\}$\langle$$2\bar{2}01$$\rangle$\textsuperscript{$-$} slip reproduced from Figure 10 of \citet[][and references therein]{de_1997}. (b) Hardening from 1\% to 2\% axial strain in each experiment in the early stage. (c) Average hardening in each percent strain between 2\% and 7\% axial strain in the late stage. }
        \label{fig:comparison_mech_data}
    \end{figure}

In Figure \ref{fig:comparison_mech_data}, we compare several characteristics of the mechanical data from our experiments to those from previous studies that employed similar experimental conditions \citep{Schmid_1980, Fredrich_1989, Rybacki_2021, Harbord_2023}. These characteristics are the differential stress at 1\% axial strain ($\sigma$\textsubscript{0.01}), the amount of hardening from 1\% to 2\% axial strain ($\sigma\textsubscript{0.02}-\sigma\textsubscript{0.01}$), and average hardening per percent axial strain within the range 2\textendash{}7\% axial strain (($\sigma\textsubscript{0.07}-\sigma\textsubscript{0.02})/5$). Our measurements of $\sigma$\textsubscript{0.01} and the strain hardening features are consistent with the results of these previous studies. This similarity suggests that our experiments and associated interpretations are broadly representative of the behaviour of calcite rocks under similar experimental conditions.\par

To analyse the inelastic behaviours of marble undergoing semi-brittle deformation, we consider the evolution of stress from soon after the yield point through subsequent strain hardening (Figure \ref{fig:strain_series_data}a). In particular, we separate the strain-hardening behaviour into two stages on the basis of the variation in the rates of hardening and microstructural evolution. From 0\% to 2\% axial strain, which we term the early stage, the hardening modulus from the three sets of experiments drops sharply with axial strain (Figure \ref{fig:strain_series_data}b). During this early stage, the evolution of twin density, true twin spacing and thickness, mean GOS, GND density, and fracture intensity is the same within uncertainty among the three sets of experiments conducted at different temperatures (Figure \ref{fig:twin_data}\textendash{}Figure \ref{fig:fracture_data}). During this stage, axial strain is accommodated almost entirely by mechanical twinning (Figure \ref{fig:twin_data}d). The true twin spacing sharply decreases with increasing strain (Figure \ref{fig:twin_data}b), and GND density remains low in this early stage (Figure \ref{fig:GOS_GND_data}b). \par

Beyond 2\% strain, which we term the late stage, the hardening modulus decreases modestly with further strain (Figure \ref{fig:strain_series_data}b). In the late stage, the samples from all three sets of experiments start to develop more lattice curvature, observed as substantial increases in estimated GND density (Figure \ref{fig:GOS_GND_data}b), whereas true twin spacing decreases more modestly with axial strain than in the early stage (Figure \ref{fig:twin_data}b). In particular, in the late stage, the samples deformed at room temperature have greater twin density (Figure \ref{fig:twin_data}a), lesser GND density (Figure \ref{fig:GOS_GND_data}b), and greater fracture intensity (Figure \ref{fig:fracture_data}), relative to the samples deformed at 200\textdegree{}C and 350\textdegree{}C. In contrast, the measurements of these microstructural elements are within uncertainty of each other between the samples deformed at 200\textdegree{}C and 350\textdegree{}C (hereafter referred to as high experimental temperatures). \par

In the following subsections, our main purpose is to correlate the mechanical data with the microstructural data in more detail. First, in Section \ref{subsec: Development of semi-brittle deformation at macroscopic scale}, we discuss the microstructural evolution and how the associated deformation mechanisms accommodate semi-brittle deformation. Following this, we discuss the relationships between microstructures and macroscopic stress in Sections \ref{subsec: Stress differential at room and high experimental temperatures}\textendash{}\ref{subsec: A phenomenological model for ductile semi-brittle deformation of marble}. In Section \ref{subsec: Stress differential at room and high experimental temperatures}, the stress at the onset of inelasticity is found to be mainly determined by the critical resolved shear stress for dislocation glide. Then, the potential mechanisms of strain hardening are discussed in Section \ref{subsec: Strain hardening}. Following such interpretation, in Section \ref{subsec: A phenomenological model for ductile semi-brittle deformation of marble}, a phenomenological model for strain hardening is proposed with true twin spacing and GND density as state variables. \par

\subsection{Microstructural evolution during semi-brittle deformation}
\label{subsec: Development of semi-brittle deformation at macroscopic scale}
    The microstructural evolution of Carrara marble during semi-brittle deformation can be inferred from the results of our mechanical testing and microstructural characterisation, as follows. During the initial increment of loading, the randomly oriented grains undergo different elastic strains due to their anisotropic stiffness. With loading progressively increased, this elastic anisotropy causes the stress and elastic strain fields to become increasingly heterogeneous among adjacent grains. \par
    
    Once the stress locally exceeds a critical stress for twinning, dislocation glide, or microcracking, permanent deformation occurs and partially relaxes the intergranular stress heterogeneity resulted from elastic anisotropy. As the critical resolved shear stress (CRSS) to initiate twinning is low (Figure \ref{fig:comparison_mech_data}a), twinning is the primary deformation mechanism in the very early stages, even before 0.5\% strain. Because calcite only has three possible $e$ twin systems, the intergranular heterogeneity in strain cannot be fully accommodated by twins alone as, according to the von Mises criterion \citep{Mises_1928}, five independent slip/twin systems are required to allow sufficient strain components in the absence of other significant processes to accommodate strain. Also, the anisotropy of the twinning process induces further intergranular interaction. Due to these effects, stress differences between grains cannot be completely relieved by twins, so heterogeneity in elastic strain partially remains and the material strain hardens. \par
    
\begin{figure}[H]
            \centering
            \includegraphics[width=0.8\textwidth]{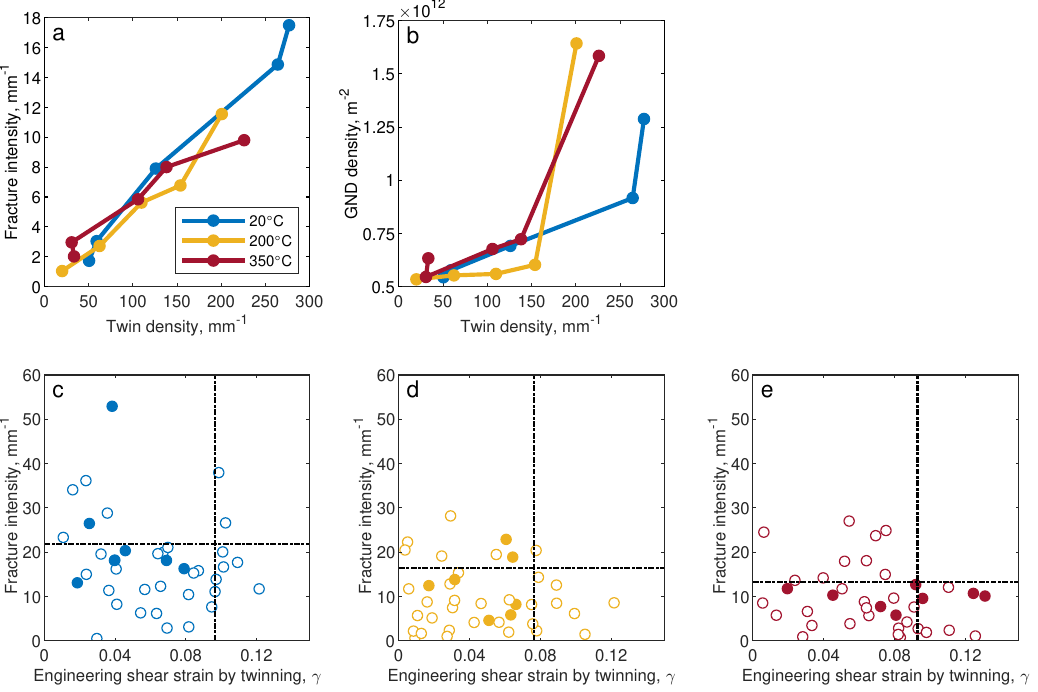}
            \caption{(a) Fracture intensity against twin density. The temperature colour code is the same among all the subplots. (b) GND density against twin density. (c–g) Fracture intensity against engineering shear strain $\gamma$ accommodated by twins (eqn. \ref{eqn: engineering_twin_strain}) of each grain selected for twin characterisation at about 7.5\% macroscopic strain. The intercepts of the dashed lines with either the horizontal or vertical axis correspond to the 80th percentile for the respective type of microstructural measurement. Markers in the top right area confined by broken lines are thus considered as grains with two simultaneously high microstructual measurements. The solid markers represent grains with GND density higher than 80th percentile, and open markers are grains with GND density lower than that.}
            \label{fig:frac_GOS}
        \end{figure}

    With further macroscopic strain, stress concentrations induced by anisotropic elasticity and twinning, exceed the critical stress to initiate intragranular microfracturing and/or dislocation glide. From our measurements, fracture intensity is linearly proportional to the increasing twin density, regardless of temperature (Figure \ref{fig:frac_GOS}a). This relationship suggests that intergranular stress heterogeneity induced by anisotropic twinning is possibly the main driver of fracturing. Likewise, as the stress rises during strain hardening, dislocation glide will be activated in progressively increasing volume fractions of the material. However, the relation between GND density and twin density is nonlinear (Figure \ref{fig:frac_GOS}b). GND density increases at a low rate in the early stage but develops rapidly in the late stage. This is possibly because, unlike how microfractures and twins must inevitably be produced by the associated processes of cracking and twinning respectively, individual dislocations have potential to glide without leaving lattice curvature if they are removed from the crystal at the end of their path. Only dislocations that glide part way through a grain and remain in the crystal (i.e., marking a gradient in plastic strain) impart remaining lattice curvature. As such, low GND densities in the early stage do not preclude some dislocation activity. \par

    In the late stage, during which twins, intragranular fractures and GND density have adequately developed, no grains simultaneously have  high GND density, high fracture intensity and high engineering shear strain due to twinning (Figure \ref{fig:frac_GOS}c\textendash{}e). Among the 41\textendash{}44 grains selected for each measurement, only a few grains have high quantities in any two measurements at each temperature. In most cases, those grains with high GND density ($>$ 80th percentile) do not have high fracture intensity or high twinning strain. The low probability for co-occurrence of any two high values of these microstructural quantities (Figure \ref{fig:frac_GOS}c\textendash{}e) suggests that the deformation in some of the measured grains may be selectively dominated by a single deformation mechanism (i.e., fracturing, dislocation glide, or twinning). Most grains underwent moderate levels of two or three microstructural activities to accommodate local strain. \par

    It should be noted that the relative displacement between grains was not possible to evaluate with our approach. Complementarily, microscale strain maps from deformed Carrara marble as split cylinders revealed that local strains along grain boundaries are associated with twinned grains at low temperatures (400\textdegree{}C and 500\textdegree{}C)\citep[Section 3.5 in][]{Quintanilla-Terminel_2016}. This observation suggests that grain-boundary sliding at low temperatures must occur, but not as a pervasive process to accommodate strain. Instead, the relative displacement/rotation between grains at low temperatures is specifically related to intergranular interaction induced by twinning. \par

\subsection{Stress differential between room temperature and high experimental temperatures}
\label{subsec: Stress differential at room and high experimental temperatures}
    The previous subsection discussed the process of accommodating strain during semi-brittle deformation of marble. The deformation mechanisms associated with the three key microstructures (i.e., dislocations, twins, and microfractures) also determine the strength of the samples at the onset of inelasticity and influence the strength evolution through the process of strain hardening. Therefore, in the following subsections, we discuss the relationships between microstructures and macroscopic stress. \par

    An obvious feature of the stress data is that the strength of Carrara marble at any given finite is substantially greater at room temperature than at 200\textdegree{}C and 350\textdegree{}C (Figure \ref{fig:comparison_mech_data}a). This stress differential appears at the onset of inelastic deformation and remains almost constant, around 120 MPa, up to 8\% axial strain. Similar measurements have been obtained under similar confining pressures of 300\textendash{}400 MPa in the previous studies compiled in Figure \ref{fig:comparison_mech_data}a \citep{Schmid_1980, Fredrich_1989, Rybacki_2021, Harbord_2023}. As there are no substantial differences in microstructure between samples deformed in the early stage at the three experimental temperatures, the strength difference between the measurements at room temperature and high experimental temperatures cannot be attributed to a shift in the relative activity of the three major deformation mechanisms (i.e., twinning, fracturing, dislocation glide), but rather to the temperature sensitivity of the strength-limiting mechanism(s). Given that fracture toughness is mostly insensitive to temperature \citep[e.g.,][]{Al_2000, Chandler_2017}, the contribution of fractures to the stress differential is likely negligible. Also, the critical resolved shear stress of $e$ twinning in single crystals is as low as 10 MPa at room temperature and decreases to about 5 MPa at high experimental temperatures \citep{de_1997}, and therefore this weak temperature dependence is also insufficient to account for the stress differential. Accordingly, the temperature dependence of the onset of inelasiticity can only be attributed to dislocation activity. The critical resolved shear stresses (CRSSs) for slip systems in single crystals are compiled in Figure 10 of \citet{de_1997} and reproduced in Figure \ref{fig:comparison_mech_data}a. The CRSS of the easy slip system, $r$ slip \{10$\bar{1}$4\}$\langle$$\bar{2}021$$\rangle$, exhibits a nonlinear dependence on temperature, decreasing from approximately 150 MPa at room temperature to about 50 MPa at 200\textdegree{}C, but only decreasing to about 20 MPa at 350\textdegree{}C. As such, the temperature dependencies of the strengths of single crystals and Carrara marble are similar Figure \ref{fig:comparison_mech_data}a, whilst the absolute strengths vary due additional factors, such as to differences in average resolved shear stress (i.e., due to differences in average Schmid factor) and differences in grain size \citep{Harbord_2023}. \par

    The inference that stress at the onset of inelastic deformation is controlled by dislocation glide needs to be reconciled with the observation that, in calcite aggregates, twinning is the primary mechanism to accommodate axial strain in the early stage (Figure \ref{fig:twin_data}d). However, twinning, as inhomogeneous deformation at grain scale (Figure \ref{fig:microstructural_data_processing}d), would induce intergranular stress heterogeneity. For calcite grains oriented unfavourably for twinning and fracturing, the accumulated intergranular stress would induce $r$ and $f$ slip. Whilst dislocation glide is not the dominant strain accommodating mechanism during the early stage, dislocation activity can be inferred from orientation maps and FSE images. For example, local lattice curvature is present in samples deformed to a strain of 1\% (Figure \ref{fig:marble_image}). These grains with lattice curvature are possibly stronger than other grains permanently deformed by fracturing or twinning, and therefore dislocation glide in these grains may have controlled the macroscopic strength of the samples. \par
        
\subsection{Strain hardening}
\label{subsec: Strain hardening}
    
    \subsubsection{Strain hardening by hindering dislocation glide}
    \label{subsubsec: Strain hardening by hindering dislocation glide}
\begin{figure}[H]
        \centering
        \includegraphics[width=1\textwidth]{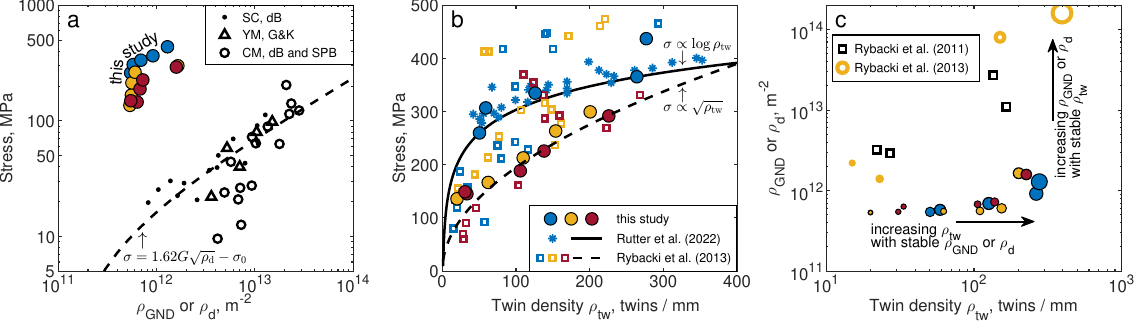}
        \caption{(a) Differential stress against geometrically necessary dislocation density from this study and total dislocation density of calcite single crystals (SC), Yule marble (YM), and Carrara marble (CM) from previous studies. The data and fitting dashed line from \citet{de_1996} (dB) were obtained from uniaxial deformation of single-crystal calcite in the [40$\bar{4}$1] direction at temperatures of 550\textendash{}800\textdegree{}C. $\sigma_0$ in the function of the dashed line is a stress constant without definitive theory foundation \citep{de_1996}. The Yule marble was deformed by \citet{Heard_1972} at temperatures of 500\textendash{}800\textdegree{}C followed by microstructural characterisation by \citet{Goetze_1977}(G\&K). The Carrara marble was deformed at temperatures of 600\textendash{}1050\textdegree{}C and examined by \citet{Schmid_1980} (SPB) and \citet{de_1996}. (b) Differential stress against twin density from deformation experiments on Carrara marble in similar temperature ranges. Blue, yellow and red respectively represent temperature ranges of 20\textendash{}100\textdegree{}C, 200\textendash{}250\textdegree{}C, and 300\textendash{}400\textdegree{}C. Data of \citet{Rybacki_2013} are from confining pressures of 100\textendash{}400 MPa and strain rates of 1$\times$10\textsuperscript{-3} s\textsuperscript{-1} to 1$\times$10\textsuperscript{-6} s\textsuperscript{-1} (including samples CM22, CM25, CM34, CM40, CM41, and CM46 in compression, and samples CMDB1, CMDB6, and CMDB8 in torsion with shear stress converted to equivalent differential stress). The broken line is from eqn. 6 of \citet{Rybacki_2013}. The solid line is eqn. 2 of \cite{Rutter_2022} fitted to data from a dog-bone shaped sample deformed at a confining pressure of 225 MPa and at a nominal axial strain rate of 4$\times$10\textsuperscript{-5} s\textsuperscript{-1}.  (c) $\rho_\mathrm{GND}$ or $\rho_{\mathrm{d}}$ against twin density. Colour-filled circles are from this study. Black squares are from calcite veins in drill core from the San Andreas Fault \citep{Rybacki_2011}. Yellow open circle are from CMDB5 in \citet{Rybacki_2013}, deformed in torsion at a temperature of 150\textdegree{}C and confining pressure of 400 MPa. The sizes of all the circles are proportionate to the magnitude of equivalent differential stress.}
        \label{fig:comparison_hardening_mechanism}
    \end{figure}

     Section \ref{subsec: Stress differential at room and high experimental temperatures} discussed the importance of dislocation glide in controlling the onset of inelasticity, and here we consider its role in the subsequent strength evolution. Dislocation glide can be hindered by interactions between dislocations and obstacles (e.g., other dislocations, twin boundaries) that can increase in intensity with macroscopic strain \citep{Taylor_1934mechanism}. Furthermore, the expansion of a dislocation loop is resisted by line tension as it emanates from between the two pinning points of a Frank-Read source \citep[][p.125]{Weertman_1971}. These effects generate an internal stress $\sigma_{\mathrm{i}}$ that hinders dislocation glide and generally has the form
    \begin{equation}
        \sigma_{\mathrm{i}} \propto \frac{Gb}{\lambda},  
        \label{eqn:Taylor}
    \end{equation}
    where $G$ is shear modulus, $b$ is the magnitude of Burgers vector, and $\lambda$ is the average spacing of dislocations or their pinning points. With plastic strain, $\lambda$ is expected to decrease due to increasing dislocation density \citep{Kocks_2003}. Therefore, in studies on mechanical properties of metals \citep[e.g.,][]{Mecking_1981, Bouaziz_2001} and silicate minerals, such as olivine, quartz, and plagioclase feldspar \citep{Thom_2022, Breithaupt_2023}, $\lambda$ is usually approximated as the square root of dislocation density ($\rho_{\mathrm{d}}$), following 
        \begin{equation}
            \sigma_{\mathrm{i}} = \alpha{}Gb\sqrt{\rho_{\mathrm{d}}},
            \label{eqn:Taylor relation}
        \end{equation}
    known as the Taylor equation, where $\alpha$ is a geometric coefficient  \citep{Taylor_1934mechanism}. Calcite single crystals also exhibit this relationship, as demonstrated by the $\sigma$-$\rho_{\mathrm{d}}$ data from \citet{de_1996} presented in Figure \ref{fig:comparison_hardening_mechanism}a. These mechanical data were obtained by uniaxial compression in the [40$\bar{4}$1] direction at temperatures of 550\textendash{}800\textdegree{}C and ambient pressure. $\rho_{\mathrm{d}}$ was directly measured as the number of dislocations per unit area by transmission electron microscopy. \par

    To explore the effect of increasing dislocation density during strain hardening in our samples, Figure \ref{fig:comparison_hardening_mechanism}a presents stress against GND density ($\rho_\mathrm{GND}$). We note that apparent GND densities are inversely proportional to the length scale over which they are measured for two reasons \citep{wallis_2016}. First, the noise floor in estimates of GND density decreases with increasing length scale as the effect of noise in the lattice rotations becomes proportionately less. Second, a greater fraction of the total dislocation density will become statistically stored dislocations with increasing length scale as the curvature produced by each dislocation becomes more likely to be cancelled by a dislocation with a Burgers vector of opposite sign. As our estimates of $\rho_{\mathrm{GND}}$ were computed over the length scale of the grain size (i.e., the largest meaningful length scale) they likely provide a lower bound on the total dislocation density in each sample. A comparison between the GND densities from our study and the total dislocation densities in single crystals measured by \citet{de_1996} reveals that our samples deformed to high stresses (i.e., the four highest stresses in the series deformed at room temperature and the two highest stresses in each of the series deformed at higher temperatures) exhibit a similar proportionality between $\sigma$ and $\sqrt{\rho_\mathrm{GND}}$ as do the single crystals (i.e., both datasets exhibit similar slopes in Figure \ref{fig:comparison_hardening_mechanism}a). This similarity suggests an important role of dislocation interactions in hindering dislocation glide during strain hardening in the late stage. \par 

    Our samples deformed to low final stresses (i.e., the two lowest stresses in the series deformed at room temperature and the four lowest stresses in each of the series deformed at higher temperatures) exhibit a different relationship between $\sigma$ and $\rho_\mathrm{GND}$, whereby stress increases with little change in $\rho_\mathrm{GND}$ (Figure \ref{fig:comparison_hardening_mechanism}a). Again, this effect is mirrored in tests on coarse-grained marble deformed to steady state at temperatures greater than 500\textdegree{}C (Figure \ref{fig:comparison_hardening_mechanism}a). In these previous data, $\sigma$ increases rapidly with relatively little change in $\rho_\mathrm{d}$ at low stress and only follows $\sigma \propto \sqrt{\rho_\mathrm{d}}$ at relatively high stresses (Figure \ref{fig:comparison_hardening_mechanism}a) \citep{Goetze_1977, Schmid_1980, de_1996}. After evaluating the influence from recovery kinetics and experimental unloading/cooling effects, \citet{de_1996} proposed that this effect arises due to strain incompatibility at grain boundaries in polycrystals, which is absent in single crystals. The strain incompatibility between grains would produce GND near grain boundaries to maintain compatible deformation of the bulk sample. Based on the assumption that each GND contributes less hardening than does a SSD, \citet{de_1996} proposed a non-linear relation between stress and $\sqrt{\rho_{\mathrm{d}}}$ for calcite polycrystals (his eqns. 12 and 17). However, the designation of a dislocation as a GND or SSD is an observational description based on whether or not it generates net lattice curvature over the length scale of interest defined by the observer and does not in itself denote a fundamental change in the physical properties or associated processes of the dislocation, at least not without further assumptions. As such, the propensity for a dislocation to generate hardening by either short-range interactions (e.g., junction formation) or long-range elastic interactions depends not on the designation of the dislocation set by the observer, but rather on the details of the positions, geometries, and types of nearby dislocations. For example, groups of dislocations of the same sign generated near grain boundaries could efficiently generate hardening if they are in a geometry that allows them to interact with other dislocations generated in the grain interior, such as being on glide planes of the same slip system. \par

    Alternatively, increases in stress due to grain boundaries and twin boundaries hindering dislocation glide is a potential explanation for the difference between single crystals and polycrystals at low stresses in Figure \ref{fig:comparison_hardening_mechanism}a. In de Bresser's experiments, the calcite single crystals were uniaxially compressed along [40$\bar{4}$1], which leads to Schmid factors for $e$ twins of 0 or 0.12, much lower than the Schmid factors for $r$ and $f$ slip \citep[Table 1 in][]{de_1990}, while the CRSSs for $e$ twinning, and $r$ and $f$ slip are similar at high temperature $>$ 500\textdegree{}C (Figure 10 of \cite{de_1997}). Therefore, twins are less abundant in these single crystals (e.g., Figure 6a in \citet{de_1990}) than in marble (e.g., Figure 6a–c in \citet{de_2005}; Figure 8g–h in \citet{Rybacki_2021}). Accordingly, twin density may serve as a secondary variable that reduces the apparent sensitivity of the macroscopic stress to $\rho_{\mathrm{d}}$ in the low stress regime (Figure \ref{fig:comparison_hardening_mechanism}a).  \par
    
    Figure \ref{fig:comparison_hardening_mechanism}b presents stress against twin density $\rho_{\mathrm{tw}}$ from our samples and previous studies conducted on Carrara marble at low temperatures similar to those of our study. Twin density linearly increases with stress in our samples. An empirical relation that $\sigma \propto \mathrm{log}(\rho_{\mathrm{tw}})$ was proposed in \citet{Rutter_2022}, while similarly, \citet{Rybacki_2013} suggested the relation as $\sigma \propto \sqrt{\rho_{\mathrm{tw}}}$. Although the two proposed relations cannot be distinguished from the scattered measurements (Figure \ref{fig:comparison_hardening_mechanism}b), a correlation between stress and twin density has been generally accepted \citep[e.g.,][]{Lacombe_1992, Amrouch_2010, Seybold_2023}. \par
    
    Furthermore, microstructural observations (i.e., EBSD, high-angular resolution EBSD, and FSE data) indicate that some twin boundaries in Carrara marble do impede dislocation glide in instances where they have low transmission factors based on the orientation relationships between the applied stress, slip system, and twin system \citep{Harbord_2023}. Thus, for our measurements, the increase of stress in the early stage with little increase of $\rho_{\mathrm{GND}}$ is plausibly attributable to an increase in the spatial density of twin boundaries, which hinder dislocation glide, similar to the previous observations made at steady state (Figure \ref{fig:comparison_hardening_mechanism}a) \citep{Rybacki_2021}. \par

    Overall, the relationships among our microstructural and mechanical data suggest that the strain hardening in marble is likely to be controlled by increasing resistance to dislocation glide, initially with twin boundaries as the main barriers that are increasing in abundance, progressively complemented by dislocation structures. In the early stage, we observed increasing $\rho_{\mathrm{tw}}$ with stable $\rho_\mathrm{GND}$. In the late stage, $\rho_{\mathrm{tw}}$ tends to become stable while $\rho_\mathrm{GND}$ significantly increases. The measurements of $\rho_{\mathrm{tw}}$ and $\rho_{\mathrm{d}}$ on the same samples from calcite veins in natural faults \citep{Rybacki_2011} and experiments \citep[CMDB5 in][]{Rybacki_2013} generally follow this pattern (indicated by arrows in Figure \ref{fig:comparison_hardening_mechanism}c). \par
    
    \subsubsection{Frictional slip}
        \label{subsubsec: Frictional slip}
\begin{figure}[H]
            \centering
            \includegraphics[width=0.9\textwidth]{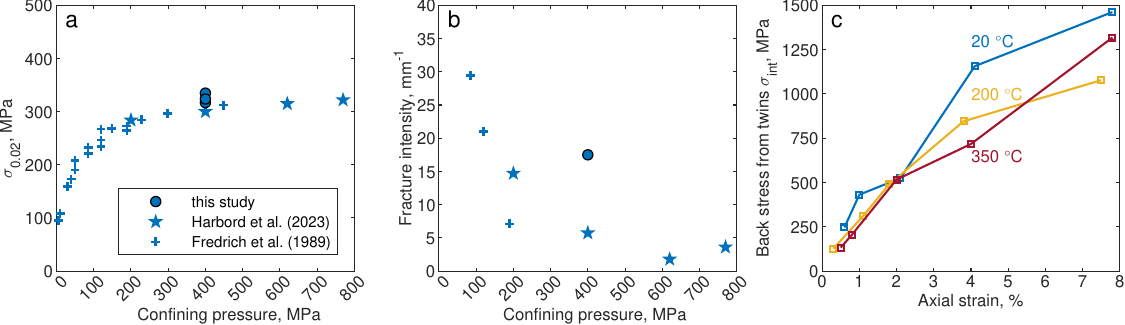}
            \caption{(a) Differential stress at 2\% axial strain ($\sigma$\textsubscript{0.02}) for Carrara marble deformed at room temperature and different pressures. (b) Fracture intensity of Carrara marble deformed at room temperature and different confining pressures. The data from \citet{Fredrich_1989} are from samples deformed to strains of 4.6\textendash{}5.5\%. The data from this study and \citet{Harbord_2023} are from samples deformed to strains of about 7.5\%. The data point from this study was acquired by fracture tracing, different from the line-intercept methods of \citet{Fredrich_1989} and \citet{Harbord_2023}. (c) Intergranular back stress from twins from eqn. (\ref{eqn:apparent twin back stress}).}
            \label{fig:stress_pressure_back stress}
        \end{figure}

        Plausibly, the strain-hardening behaviour might also be attributable to frictional slip on shear cracks during compressive loading under confinement. At sufficiently high pressures, any pre-existing cracks are closed and initial loading induces a linear elastic response determined by the intrinsic Young's modulus of the intact solid. As stress and strain increase, the behaviour deviates from linearity (e.g., Figure \ref{fig:data_processing}b). If crystal-plastic behaviour were not involved, the non-linearity could be explained by frictional displacement on pre-existing microfractures and observed as apparent strain hardening after yielding \citep[e.g.,][]{Walsh_1965, David_2020}. However, in experiments at confining pressures of 400 MPa and above, the macroscopic strength of marble (Figure \ref{fig:stress_pressure_back stress}a) and the fracture intensity of samples (Figure \ref{fig:stress_pressure_back stress}b) \citep{Fredrich_1989, Harbord_2023} exhibit minimal pressure dependence. Also, the intragranular fractures usually do not propagate across an entire grain \citep[][also Figure \ref{fig:marble_image} of this study]{Fredrich_1989}. These observations suggest that frictional slip of fractures is mostly limited in our samples, and thus cannot contribute much to the strain-hardening behaviour.\par

    \subsubsection{Intergranular back stress at twin tips}
        \label{subsubsec: Intergranular back stress at twin tips}
        The pervasive twinning, as an heterogeneous deformation process at the grain scale (Figure \ref{fig:microstructural_data_processing}c), induces geometric complications at grain boundaries \citep[e.g.,][]{Burkhard_1993}. If a twinned grain is between two elastic grains, any intergranular back stress ($\sigma_{\mathrm{int}}$), induced by strain incompatibility at twin tips, might also contribute to strain hardening. The calculation of $\sigma_{\mathrm{int}}$ has been proposed by \citet{Mitchell_1991} as
        \begin{equation}
                \sigma_{\mathrm{int}} = \frac{y_0\varepsilon_{\mathrm{t0}}G_{\mathrm{VRH}}}{d_\mathrm{{twsp}}},
            \label{eqn:apparent twin back stress}
        \end{equation}
        where $y_0$ is the halved true width of twins, $d_\mathrm{{twsp}}$ is the true twin spacing and $\varepsilon_{\mathrm{t0}}$ is the engineering twin strain for a single twin. The stress computed by eqn. (\ref{eqn:apparent twin back stress}) is an upper bound as this equation assumes linear-elastic behaviour of the material surrounding the twins. \par
        
        The resulting estimates of $\sigma_{\mathrm{int}}$ are similar for all tested temperatures (Figure \ref{fig:stress_pressure_back stress}c). The stress estimates are least at around 250 MPa at 0.5\% strain, and increase up to 1\textendash{}1.5 GPa at 7.5\textendash{}8\% strain. Such high estimated stresses indicate that twinning in one grain leads to strong stress concentrations in adjacent grains. However, once the intergranular stress is beyond a critical stress, the stress concentrations are prevented from increasing further by intragranular flow (i.e., twinning or dislocation glide) or fracturing in the neighbouring grains. Accordingly, at the grain scale, the development of such stress concentrations at twin tips is important to the microstructural evolution. However, the stress at twin tips cannot surpass the critical stress for intragranular flow. Thus the intragranular mechanisms, rather than intergranular back stress at twin tips, dominate the macroscopic hardening behaviour. \par
    
    \subsubsection{Other possible hardening or softening mechanisms}
        \label{subsubsec: Other possible hardening mechanisms}
        Alongside dislocation glide hindered by increasing obstacles and limited frictional sliding on shear cracks, many other intragranular processes might also have some influence on the strength of deformed marble. The CRSS to induce twins is increased by intersection between different twin sets in fcc alloys \citep{Alkan_2018, Bonisch_2018}. Furthermore, the strength of marble undergoing semi-brittle deformation can be influenced by more complicated direct interaction between fractures and dislocations. Dislocations can be generated at stress concentrations at fracture tips \citep[e.g.,][]{Anderson_1986}, which facilitates plastic deformation at relatively low macroscopic stress and is, in this respect, a softening process. However, any resulting increase in the intensity of dislocation interactions or shielding of stress concentrations at the fracture tips could also contribute a hardening effect \citep[Section 7.3 in][]{Lawn_1993}. Also, the strength of tested samples may be indirectly influenced by dynamic recovery processes, including cross slip \citep{de_2002}, annihilation \citep{Nes_1997}, or dislocation removal at new free surfaces created by fracturing \citep[Chpater 9 in][]{Caillard_2003, Brantut_2024}. \par
        
        The aforementioned mechanisms in this subsection are difficult to test, because they do not have a resolvable mechanical or microstructural signature that we can identify. Also, the hardening mechanisms introduced in Section \ref{subsubsec: Frictional slip} and \ref{subsubsec: Intergranular back stress at twin tips} (i.e., frictional slip and back stress at twin tips) are either limited at high pressure or not yet clearly demonstrated to be effective hardening/softening processes in marble. Accordingly, it is most likely that  dislocation glide hindered by increasing obstacles dominates the observed strain hardening in marble. This interpretation is supported by the results of mechanical testing and microstructural observations. First, the stress differential between room temperature and high experimental temperature, which is first apparent at the onset of inelasticity and is interpreted to be associated with dislocation glide, remains nearly constant with strain (Section \ref{subsec: Stress differential at room and high experimental temperatures}). Second, the temperature independence of hardening rate is a known characteristic of hardening by dislocation interactions \citep{Hansen_2019, Kocks_2003}. Third, the relationship between the development of twin density and that of GND density with increasing strain and macroscopic stress is consistent with them playing a coupled role in controlling the strength evolution (Section \ref{subsubsec: Strain hardening by hindering dislocation glide}). \par

\subsection{A phenomenological model for ductile semi-brittle deformation of marble}  
\label{subsec: A phenomenological model for ductile semi-brittle deformation of marble}
        We develop a phenomenological model based on the assumption that strain hardening is attributed to dislocation glide hindered by twin boundaries and dislocation interactions. As indicated by eqn. (\ref{eqn:Taylor}), the shortening of length scale $\lambda$ for dislocation glide is key in this process. Following the discussion in Section \ref{subsubsec: Strain hardening by hindering dislocation glide}, interactions of dislocations with other dislocations and twin boundaries (and grain boundaries) can jointly contribute to obstruction of dislocation glide. As the interaction among dislocations and interaction between dislocations and twin/grain boundaries cannot be physically isolated, $\lambda$ in eqn. (\ref{eqn:Taylor}) is approximated as the harmonic average of the distances between dislocations and the distances between twin boundaries \citep{Breithaupt_2023}. Thus, the Taylor relation can be extended as
        \begin{equation}
            \sigma = k_\mathrm{C}\tau_{\mathrm{C0}} + \frac{k_{\mathrm{tw}}Gb}{d_\mathrm{{twsp}}} + k_{\mathrm{GND}}Gb\sqrt{\rho_{\mathrm{GND}}}
        \label{eqn:extended Taylor relation}
        \end{equation}
        where $k_\mathrm{C}$, $k_{\mathrm{tw}}$ and $k_{\mathrm{GND}}$ are three coefficients, $\tau_{\mathrm{C0}}$ is CRSS for dislocation glide, $d_\mathrm{{twsp}}$ is true twin spacing, and $\rho_\mathrm{GND}$ is GND density. In eqn. (\ref{eqn:extended Taylor relation}), the first term represents the yield stress of marble samples at different temperatures. This yield stress is influenced by the CRSS of the relevant slip system, the average Schmid factor of that slip system, and grain size \citep[e.g.,][]{Walker_1990, Harbord_2023}, which are considered to be invariable with macroscopic strain for our samples. The coefficient $k_\mathrm{C}$ and CRSS thus only describe the temperature dependence of yield stress in our application (discussed in Section \ref{subsec: Stress differential at room and high experimental temperatures}). Also, dislocation density in the Taylor equation is replaced with GND density in eqn. (\ref{eqn:extended Taylor relation}) to model our data. Thus, here $k_{\mathrm{GND}}$ is not exactly equivalent $\alpha$ in the Taylor equation (eqn. (\ref{eqn:Taylor relation})). \par

\begin{figure}[H]
            \centering
            \includegraphics[width=0.8\textwidth]{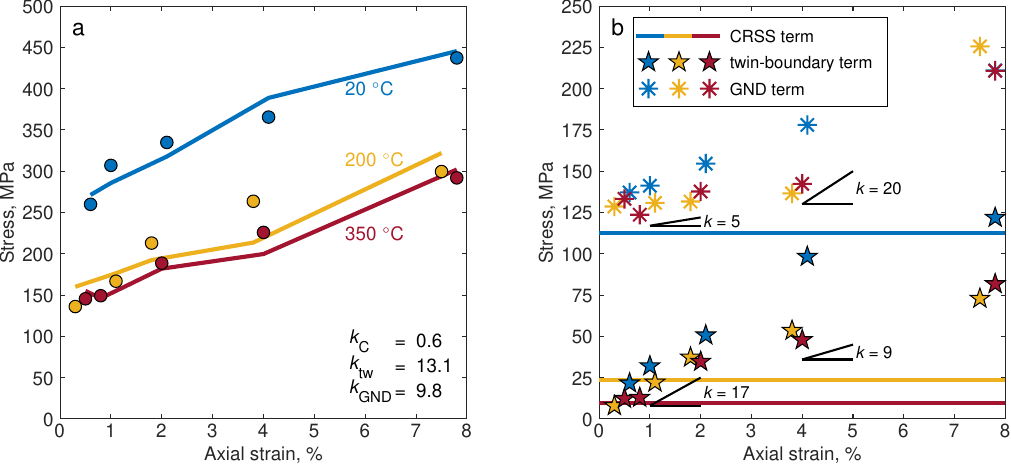}
            \caption{(a) Linear least-square inversion results (lines) with experimental measurements of stress for the fifteen samples. (b) Contribution of the four terms in eqn. (\ref{eqn:extended Taylor relation}), the extended Taylor relation, to the modelled strength. The datum for the GND term at the greatest strain at room temperature is overlapped by the datum for 350\textdegree{}C. $k$ indicates the slope of stress-strain data for twin-boundary term or GND term in the early or late stage at high experimental temperatures.}
        \label{fig:inversion}
        \end{figure}

        The three coefficients, displayed in Figure \ref{fig:inversion}a, were found by linear least-square methods applied to all the mechanical and microstructural data of this study. In this process, $\tau_{\mathrm{C0}}$ is fixed at 196.5 MPa at room temperature, 41.0 MPa at 200\textdegree{}C and 16.7 MPa at 350\textdegree{}C by the best fit line for $r$ slip in Figure 10 of \citet{de_1997}. $\frac{1}{d_\mathrm{{twsp}}}$ and $\sqrt{\rho_{\mathrm{GND}}}$ are from microstructural characterisation of the fifteen tested samples. The inversion result suggests that for unit length of state variables (i.e., twin spacing and distance between GND), the capability of increasing twin boundaries to obstruct dislocation glide is relatively greater than that of dislocation-dislocation interactions (i.e., $k_{\mathrm{tw}} > k_{\mathrm{GND}}$). Given that GND significantly outnumber twins, the absolute strength is mostly contributed by GND.\par
        
        In terms of hardening, the twins and GND play roughly equal roles, with both contributing about 100 MPa of hardening over the 8\% strain. More specifically, at room temperature, the amount of hardening increases linearly with strain for either of the two terms. At high experimental temperatures, most of the strain hardening is attributable to the twin-boundary term in the early stage, while the GND term only has weak dependence on strain (slopes indicated in Figure \ref{fig:inversion}b). In contrast, GND term contributes to most of the hardening in the late stage at high experimental temperatures, while twin-boundary term increases in a slower rate. \par
        
        In this model for semi-brittle deformation at high pressure, the role of fractures is limited to accommodation of strain incompatibility. Whether the macroscopic stress is directly influenced by fracturing remains to an open question but could be explored by another strain-series study at progressively lower pressure. The current model, based on an extended Taylor equation, serves as the baseline behaviour with the assumption that only dislocations and twin boundaries have direct effects on the strength evolution of marble. \par

        Towards a microsphysical model, it is necessary to derive the process of microstructural development as a function of macroscopic strain. With the assumption that plastic strain is completely accommodated by dislocations, the development of dislocation density could be modelled as 
        \begin{equation}
            \frac{\mathrm{d}\rho_\mathrm{d}}{\mathrm{d}\varepsilon} = \frac{1}{b\lambda} - f\rho_\mathrm{d}, 
        \label{eqn:drho_d/depsilon}
        \end{equation}
        where $f\rho_\mathrm{d}$ is a term for dislocation removal \citep{Kocks_1965, Kocks_2003, Rybacki_2021}. However, our microstructural observations suggest that dislocations accommodate only a fraction of the plastic strain. Thus, any application of Equation \ref{eqn:drho_d/depsilon} to semi-brittle deformation would require its extension to account for the comprehensive development of twinning, dislocation activity and microfracturing \citep[e.g.,][]{Brantut_2024}. Besides, the interactions between any two mechanisms could also influence the development of twin spacing and dislocation density. Such interaction between any of the two mechanisms cannot be clearly captured by the conventional EBSD mapping in this study. Further microstructural characterisation by high-angular resolution EBSD and transmission electron microscopy may better resolve this issue. Alternatively, recent progress in correlating acoustic emission waveform patterns with microstructural activities may provide a new possibility for resolving more in-situ microstructural information \citep{O’Ghaffari_2023}. From this study, the quantitative constraints on the key microstructures provide a starting point for further microstructural investigations and microphysical modelling. \par

\section{Conclusions}
In this study, three sets of marble samples were shortened to varying strains up to 8\% at experimental conditions that induce semi-brittle deformation, specifically a confining pressure of 400 MPa and temperatures of 20\textdegree{}C, 200\textdegree{}C, or 350\textdegree{}C. Twins, lattice curvature and intragranular microfractures were quantitatively examined by forescattered electron imaging and electron backscatter diffraction in the scanning electron microscope. The results reveal that, in the early stage of deformation (strain $\leq$ 2\%), deformation is primarily accommodated by twins. Lattice distortion, linked to geometrically necessary dislocations, starts to become pronounced in the later stage (strain $>$ 2\%). Intragranular fracture intensity exhibits an almost linear correlation with strain in the early stage and tends to stabilise in the late stage. Despite some nuanced variations, the overarching development of each microstructural element is similar across the different temperatures. \par

From mechanical testing, we found that the stress at the onset of inelasticity at room temperature is significantly greater than those at 200\textdegree{}C and 350\textdegree{}C. This temperature dependence of the macroscopic stress is consistent with the temperature dependence of the critical resolved shear stress for $r$ slip in calcite. Regarding strain hardening, we found that hindering of dislocation glide by decreasing twin spacing and increasing dislocation density is the dominant mechanism. Based on this interpretation, a phenomenological model for semi-brittle deformation has been proposed as the extended Taylor relation. This model contains one coefficient for the critical resolved shear stress of $r$ slip that determines the stress at the onset of inelasticity, and two coefficients that relate hardening to state variables of true twin spacing and dislocation spacing based on GND density. The microstructural data in this study provide a general overview of the evolution of the state variables with strain and experimental conditions as a precursory basis for a microphysical model of semi-brittle deformation. \par

\section*{Acknowledgements}
Emmanuel David contributed to early technical developments on the Murrell apparatus. Technical support from Harison Wiesman, John Bowles and Neil Hughes is greatly appreciated. We thank Thomas Breithaupt, Hans de Bresser, Georg Dresen and Erik Rybacki for useful discussions. This project has received funding from the European Research Council (ERC) under the European Union's Horizon 2020 research and innovation programme (grant agreement 804685/“RockDEaF” to N.B.) and from the UK Natural Environment Research Council (Grant Agreement NE/ M016471/1 to N.B.). DW acknowledges support from a UK Research and Innovation Future Leaders Fellowship (Grant Agreement MR/V021788/1).

\section*{Data availability}
The mechanical data and microstructural data (EBSD maps, node positions of traced fractures and twin information) have been uploaded to Zenodo at doi.org/10.5281/zenodo.14187374. 

\bibliography{references}

\end{document}